

\magnification=1200
\overfullrule=0pt
\parskip=5pt

\def\ket#1{|#1\rangle }
\def\bra#1{\langle #1|}
\def\bk#1{\langle #1\rangle }
\def\sbk#1{[#1]}
\def\({\left(}
\def\){\right)}
\def\[{\left[}
\def\]{\right]}

\def\b{$$\eqalignno}
\def\C{{\cal C}}
\def\h{{1\over 2}}
\def\l{\ell }
\def\p{\partial }
\def\.{\!\cdot\! }
\def\r2{\sqrt {2}}

\def\Tr{{\rm Tr}}
\def\frac#1#2{{#1\over #2}}

\pageno=0
\rightline{McGill/93-20}
\rightline{hep-ph/9308289}
\rightline{(June, 1993)}
\bigskip\bigskip\bigskip
\centerline{\bf SPINOR HELICITY TECHNIQUE AND STRING REORGANIZATION}
\medskip
\centerline{\bf FOR MULTILOOP DIAGRAMS$^\dagger$}
\bigskip\bigskip\bigskip
\centerline{C.S. Lam$^*$}
\bigskip
\centerline{Department of Physics, McGill University, 3600 University St.}
\medskip
\centerline{Montreal, P.Q., Canada H3A 2T8}
\vskip4cm

\centerline{\bf Abstract}
\bigskip \bigskip
Methods to apply the spinor helicity technique and string
reorganization to multiloop
amplitudes using the Feynman- and Schwinger-parameter representations
are reviewed and expanded.

\vfill
\leftline{$^*$ Email address: Lam@physics.mcgill.ca}
\leftline{$^\dagger$ To be published in the Canadian Journal of Physics}
\bigskip \bigskip
\eject

\centerline{\bf 1. Introduction}

Perturbation theory is taught in every text book but its actual
calculation could be very complicated. This is particularly so
for gauge theories where the presence of spin, and possibly color,
come in to tangle up the algebra. The fact that many gauge-dependent
terms occur in the intermediate steps do not help the
situation either.
Such complexity grows at an alarming rate with the number $n$ of
external particles and with the number $\l$ of loops. For example,
the number of terms in a QCD $2\to 6$ gluon process
is estimated already to be half a billion, even when no loop nor
fermion is present.
This is a serious problem that cannot be ignored because
large $n$ and $\l\not=0$
calculations are unavoidable
in high energy and high precision experiments. If polarized
cross-sections are measured helicity amplitudes are needed as well,
and this adds further to the complications.
With such an amount of complexity even large computers are powerless.
This makes it imperative to find new ways to overcome or to
reduce   computational difficulties in loops, with large
$n$, and for polarized amplitudes.

In this connection there were two important advances in the last decade.
This is the discovery of the {\it spinor helicity technique}
[1--21] and the realization of the power of
{\it superstring} calculations [22--25]. With the help
of these new techniques previously
uncalculable processes
can now be computed. For example, exact tree amplitudes for
$e^++e^-\to \mu^++\mu^-+n\gamma$ can now be
obtained provided all photons
carry the same helicity. Exact $n$-gluon tree amplitudes in the
absence of fermions are given by
the Parke-Taylor formula, which is valid
if all but at most two of the $n$ outgoing gluons have the same
helicities. These and other
tree processes were calculated using the spinor
helicity technique [1--21]. One-loop four and five polarized gluon amplitudes
have now been obtained from superstring calculations [22--25].

In its original form,
the spinor helicity technique cannot be used beyond the tree
approximation. It is also   difficult to carry out superstring
calculations beyond one loop  or to include external fermions.
These limitations have now been overcome   using parametric
representations [26,27]. As a result,
spinor helicity technique can now be applied to
any number of loops [26]. A field-theoretic substitute for superstring
calculations
is now available to allow  simplifications found in one-loop
superstring calculations to be extended to multiloop amplitudes,
with or without external fermions [27]. We shall call this field-theoretic
substitute {\it string reorganization}.

The main purpose of this paper is to review these two recent
developments. Although most of the crucial points have been made
before [26,27], there are nevertheless a number of new formulas, new
proofs, and new details added to this article.
This includes the use of spinor paths
in Sec.~4,  the new differential circuit identities
in Appendix A and the corresponding new proof of the stringlike formulas.
A technical restriction
appeared earlier [27] governing the use of the multiloop stringlike formula
has now  been removed, and stringlike formulas have also been
generalized to QCD. Local gauge invariance for internal vertices are
shown in Appendix D, a new interpretation of gauge invariance as proper-time
reparametrization invariance has been put forward. Finally,
a perturbative proof of the off-shell Gordon
identity is given in Appendix E.

In order to fulfil the dual role of a review article, that of giving
an elementary introduction to the basic ideas and  results,
but at the same
time being reasonably detailed and self-contained to be useful
to workers in the field, the article is divided into the main
text and the appendices.
The main text, from Sec.~2 to Sec.~5, contains
a discussion of  the  motivations, the problems,  and the
solutions. The proofs and other finer details are to be found
in the appendices.

The methods to be discussed in this article
are applicable to any gauge theory,
but for simplicity of presentation we shall confine ourselves
here to QED and QCD.
In the rest of this section, we shall give a brief account
of these two new developments.

The spinor helicity technique makes use of the fact that
external fermion masses can be neglected in   high energy collisions.
In that case helicity and chirality are both conserved and the
multichannel spin problem reduces to a one-channel problem
(see Sec.~4.1 for details). The
technique can be used as long as all the internal momenta are
linear combinations of the {\it massless external}
momenta, which is so for tree but not for loop amplitudes.
This is why
in its original form the technique can only be applied to tree
but not loop
diagrams. There is however nothing against {\it internal} particles
to be massive, so the technique is equally applicable for
massive $W$ and $Z$ exchanges, as well as their productions if
their subsequent decays into (nearly) massless particles are also
included in the diagram.

The main idea in the recent development [26]
is to write the loop amplitudes in the
Schwinger-parameter representation (or the related
Feynman-parameter representation) rather than the conventional
momentum-space representation.
This avoids the presence of off-shell
loop momenta and allows the
spinor helicity technique to be used (see Sec.~4). This approach
is practical because
amplitudes in parametric representations can be written down just
as easily as amplitudes in momentum representations [28--32]
(see Sec.~2).

The main avenue for carrying out  string reorganization
of multiloop amplitudes [27] is again the Schwinger-parameter
representation, though the reason is now different:
the Schwinger proper-time parameters turn out  to be just
the proper time $\tau $ of a superstring in the infinite tension
limit. It is therefore not surprising that string reorganization
takes place in the parametric rather than the momentum-space framework
(see Secs.~3 to 5).
It turns out that the use of
Schwinger representation  does not by itself give
all the string reorganizations. The use of
{\it differential circuit identities}, discussed in Appendix A, is
essential to obtaining the full result.

In this connection it is crucial to recognize the importance of electric
circuit theory (see Appendix A).
A Feynman diagram can be regarded as an electric
circuit [28--32]: the Schwinger or Feynman parameters are the resistances,
the external momenta are the currents flowing in or out of the network,
and the internal momenta are currents  along the
internal lines. The scattering amplitude in parametric representations
can be expressed in terms of
circuit quantities like power, currents,  etc.
Knowledge of the relationship between these  quantities can
 lead
to simplification of the  scattering amplitudes, as will
be seen in the rest of this article. Such knowledge includes
not only formulas contained in electrical engineering text books,
but also  differential circuit identities not normally found
in the literature.

{}From a field-theoretic point of view, string reorganization accomplishes
two things. First, it puts  internal and  spacetime variables
on a somewhat equal footing: they are all flows (current flow, spin flow,
color flow, etc.) on the network (see Secs.~3 and 4).
Secondly, it succeeds in making local gauge invariance
transparent (see Sec.~5 and Appendix D). As a byproduct of
the latter, one also sees the possibility of interpreting
gauge invariance as a reparametrization invariance in the
proper-time variable. This is discussed in Appendix D, between eqs.~(D10)
and (D11).

\bigskip
\centerline{\bf 2. Scattering Amplitudes}

The scattering amplitude for
 a Feynman diagram in $d=4-\epsilon$
 dimensions with $n$ vertices, $N$ internal
lines, and $\l$ loops is
$$A=\left[{-i\mu^\epsilon\over
(2\pi )^{d}}\right]^{\ell }\int
\prod _{a=1}^{\ell }(d^{d}k_{a}){S_{0}(q,p)\over
\prod _{r=1}^{N}(-q_{r}^{2}+m_{r}^{2}-i\epsilon )}\ ,\eqno(2.1)$$
where $k_{a}\ (1\le a\le\l$) are the loop  momenta,
$q_{r}, m_{r}\ (1\le r\le N)$
are the momenta and masses of the internal lines, and $p_{i}\ (1\le i\le
n)$ are the conserved external momenta
satisfying
$$\sum _{i=1}^{n}p_{i}=0\ .\eqno(2.2)$$
 For convenience an external momentum $p_{i}$ has
been assigned to each  vertex. If a vertex is internal, we simply have to set
the corresponding $p_{i}=0$ at the end.

The numerator function
$S_{0}(q,p)$ contains everything except the denominators of the
propagators. Specifically, it is the product
of the vertex factors, numerators of propagators,
wave functions of the external lines,
symmetry factor, and the signs associated with closed fermion loops.
All the $i$'s and $(2\pi )$'s have been included in the factor before the
integral.

The internal momenta $q_{r}$ are linear combinations of $p_{i}$ and $k_{a}$.
In the case of {\it tree} diagrams, $k_{a}$ is absent, so
$$q_{r}\equiv \sum _{i}I_{ri}p_{i}\eqno(2.3)$$
is given
by a linear combination of the external momenta alone. It is this absence of
$k_{a}$ that
enables the spinor helicity technique to be applied to tree diagrams.
See Sec.~4.
To use this technique on loop diagrams,
it is necessary to get rid of the $k_{a}$-dependence of
$q_{r}$ [26]. This can be accomplished by introducing the Feynman
parameters $\alpha_{r}$, one per internal line, so  the loop integrations can
be
explicitly carried out. The resulting Feynman-parameter representation reads
[30]
$$\eqalignno{
A=&\int [D_F\alpha]\Delta(\alpha)^{-d/2}\bar S(q,p)\Gamma\(N-
{d\l\over 2}\)\[D(\alpha,p)\]^{-N+d\l/2}\ ,&\cr
\int[D_F\alpha]&\equiv \[{\mu^\epsilon\over(4 \pi)^{d/2}}\]^\l
\(\prod_{r=1}^Nd\alpha_r\)\delta \(\sum _{r=1}^{N}\alpha_{r}-1\)\ ,&\cr
M&\equiv \sum_{r=1}^N\alpha_rm_r^2\ ,&\cr
D(\alpha ,p)&=M-P(\alpha ,p)\ ,&\cr
\bar S(q,p)&=
\sum _{k\ge 0}i^{k}{\Gamma \(N-{d\ell \over 2}-k\)
\over\Gamma\(N-{d\l\over 2}\)}
D(\alpha ,p)^k S_{k}(q,p)
\ .&(2.4)}$$
$S_{0}(q,p)$ is the same function as the one appearing in (2.1), except that
$q_{r}$ is now given by (2.3), with $k_{a}$ absent
and its dependence  replaced by the dependence of the coefficients $I_{ri}$
on the $\alpha$'s. It, along with other quantities in (2.4), can be computed
using
electric circuit theory.
See Appendix A. If the Feynman diagram is regarded as an electric
circuit, $\alpha_{r}$ as the
resistance of the $r$th internal line and $p_{i}$ as the
current flowing out of the $i$th vertex, then $q_{r}$ is the
current flowing through the $r$th line and
$P$
is the power consumed by the network.
Thus
$$P=\sum _{r=1}^{N} \alpha _{r}q_{r}^{2}=
\sum _{i,j=1}^{n}Z_{ij}(\alpha )p_{i}\.p_{j}\ ,
\eqno(2.5)$$
where $Z_{ij}(\alpha )$ is the {\it impedance matrix}.
The quantity $\Delta $ is the
discriminant appearing in all these circuit quantities.

Suppose the numerator function $S_{0}(q,p)$ is a
polynomial in $q$ of degree
$e$. Then $S_{k}(q,p)$ is defined to be
 a polynomial in $q$ of degree $e-2k$, obtained from $S_{0}(q,p)$ by
 contracting $k$ pairs of $q$'s in all possible ways  and
 summing over all the contracted results. The rule for contracting a
pair of $q$'s is:
$$q^{\mu }_{r}q^{\nu }_{s}\to -{i\over 2}H_{rs}(\alpha)g^{\mu \nu }
\equiv q^{\mu }_{r}\sqcup q^{\nu }_{s}\ .\eqno(2.6)$$

The circuit quantities in (2.4) and (2.6)
are given explicitly by the following formulas [30]:
$$\eqalignno{
\Delta (\alpha )&=\sum _{T_{1}}(\prod ^{\ell }\alpha )\ ,&(2.7)\cr
P(\alpha ,p)&=\Delta (\alpha )^{-1}\sum _{T_{2}}
(\prod ^{\ell +1}\alpha )
(\sum _{1} p)^{2}\ ,&(2.8)\cr
q_{r}(\alpha ,p)&=\pm \Delta (\alpha )^{-1}\sum _{T_{2}(r)}\alpha _{r}^{-1}
(\prod ^{\ell +1}\alpha )(\sum _{1} p)\ ,&(2.9)\cr
H_{rr}(\alpha )&=-\Delta (\alpha )^{-1}\partial \Delta (\alpha )/\partial
\alpha _{r}\ ,&(2.10)\cr
H_{rs}(\alpha )&=\pm \Delta (\alpha )^{-1}
\sum _{T_{2}(rs)}(\alpha _{r} \alpha _{s}
)^{-1}
(\prod ^{\ell +1}\alpha )\ ,\quad(r\not= s)\ .&(2.11)}$$
The meaning of these formulas are as follows.
An $\ell $-loop diagram can be made into a connected tree diagram (a `1-tree')
by cutting $\ell $ lines, and into a diagram with two disjoint trees
(a `2-tree') by cutting $\ell +1$ lines. $\Delta (\alpha )$ is given by
the sum over the set $T_{1}$ of all 1-trees so obtained, with the summand
equal to
the product of the $\alpha $'s of the cut lines. $\Delta P(\alpha ,p)$
is given by the sum over the set  $T_{2}$ of all 2-trees so obtained, with the
summand to be
the product of the $\alpha $'s of the cut lines, times the square
of the sum of all the external momenta $p_{i}$ attached to one of these two
trees. Finally, let $T_{2}(r)$ be the set of all 2-trees in which line
$r$ is cut, and such that when the line $r$ is inserted back
a 1-tree results, and $T_{2}(r,s)$ be the set of all 2-trees in  which
lines $r$ and $s$ are cut, and such that when either the line $r$ or
the line $s$ is inserted back a 1-tree results. Then
$\Delta q_{r}(\alpha ,p)$ is given by the sum of all 2-trees in $T_{2}(r)$
 with the summand equal to
the product of $\alpha $'s of all the cut lines except the $r$th, times the
sum of all the external momenta $p_{i}$ attached to one of these two trees,
and $\Delta H_{rs}(\alpha )$ is given by the sum of all 2-trees
in $T_{2}(r,s)$ with the summand equal to the product
of $\alpha $'s of all the cut lines except the $r$th and the $s$th.
The sign in (2.9) can be determined from the direction of the current
flow in an obvious way. The sign in (2.11) is $+$ or $-$ depending
on whether lines $r$ and $s$ point to the same tree, or opposite trees.

For example, using eqs.~(2.7)--(2.11), the relevant quantities for Fig.~1
can be computed to be
\b{
\Delta &=\alpha _{1}+\alpha _{2}+\alpha _{3}+\alpha _{4}\ ,&\cr
\Delta P&=\alpha _{1} \alpha _{3}(p_{1}+p_{3})^{2}+\alpha _{2}
\alpha _{4}(p_{1}-p_{2})^{2}
+\alpha _{1} \alpha _{4} p_{1}^{2}+\alpha _{1} \alpha _{2}
 p_{2}^{2}+\alpha _{3} \alpha _{4} p_{3}^{2}
 +\alpha _{2} \alpha _{3} p_{4}^{2}\ ,&\cr
\Delta q_{1}&=\alpha _{2}p_{2}+\alpha _{3}
(p_{2}+p_{4})+\alpha _{4}(p_{2}-p_{3}+p_{4})\ ,&\cr
\Delta H_{13}&=+1\ .&(2.12)}$$
Other $q_{r}$ and other $H_{rs}$ can be similarly calculated.

Instead of Feynman's method discussed above, one can use
Schwinger's proper time formalism. In that case, one writes
$$
  \frac {1}{m_{r}^{2}-q_{r}^{2}-i\epsilon }=i\int _{0}^{\infty }d\alpha
_{r}\exp[-i\alpha _{r}(m_{r}^{2}-q_{r}^{2})]\eqno(2.13)
$$
and integrates over the internal momenta [see (A19) and (A20)].
The result is [30]
$$\eqalignno{
A=&\int [D_S \alpha ]\Delta (\alpha )^{-d/2}S(q,p)\exp[-i\{M-P\}]\ ,
\qquad {\rm where}&\cr
\int [D_S \alpha ]\equiv &\[{(- i)^{d/2}\mu^\epsilon
\over  (4 \pi)^{d/2}}\]^{\l }i^{N}
\int _{0}^{\infty }(\prod _{r=1}^{N}d\alpha _{r})\ ,&\cr
S(q,p)\equiv &\sum _{k\ge 0}S_{k}(q,p)\ ,
&(2.14)}$$
with the same functions $M,\Delta ,S_{k},q_{r}$ and $P$
as those appearing in (2.4). These two parametric
representations are actually related in a simple way. If for the moment
we use $\alpha'_{r}$ to denote the variables $\alpha_{r}$ in (2.14), and define
$\sum _{r}\alpha'_{r}=\lambda , \alpha'_{r}=\lambda \alpha_{r}$, then $\sum
_{r}\alpha_{r}=1$. Carrying
out the integration over $\lambda $ in (2.14), eq.~(2.4) will be obtained.

It is useful to point out the similarity and the difference between
(2.14) and (2.1). The loop-momentum integration in (2.1) is replaced
by the Schwinger-parameter integration $\int [D_S \alpha ]
\Delta ^{-d/2}$ in (2.14); the numerator function $S_{0}(q,p)$ is replaced
by $S(q,p)=\sum _{k\ge 0}S_{k}(q,p)$, and the propagator factor
$(-q_{r}^{2}+m_{r}^{2}-i \epsilon )^{-1}$ is replaced by
$$T(q_{r},m_{r},\alpha _{r})=\exp[-i \alpha _{r}(m_{r}^{2}-q_{r}^{2})]\
,\eqno(2.15)$$
which is the time development factor for a particle with momentum $q_{r}$
propagating under the Hamiltonian $H_{op}=p_{op}^{2}-m_{r}^{2}$  over
a `proper time' $\alpha _{r}$.

As discussed in the Introduction, a gauge-theory amplitude
may contain many many terms. In the language of (2.4) and (2.14),
 this is reflected
in the complexity of computing
 $S_{k}(q,p)$
and the corresponding $\alpha $-integrals.
Methods to simplify their calculations will be the main topic for the
rest of this paper.
When $S_{k}$ and the other circuit quantities in (2.4) and (2.14) are
thus computed, there is still the remaining task of carrying out the
$\alpha_{r}$
integrations.
In a small number of cases analytical results are known, but in the
great majority of situations one would have to resort to numerical
methods at that stage.
We will not discuss the problem  further.
If renormalization is required, it has to be carried out
in the usual way as well [32,33],
and we shall not go into that aspect  either.

It would be convenient to consider all scattering processes in a standard form,
in which
all gluons and photons are outgoing and no antifermions are present.
This can be achieved by
 moving particles from one
side of the reaction to another if necessary, thus converting
antifermions into fermions and incoming momenta
 into outgoing,
 without altering the scattering amplitude $A$. We shall
assume in the rest of this article that this has already been done.

\bigskip \bigskip
\centerline{\bf 3. Color Flow}

The total scattering amplitude $A_{tot}$ of a QCD process (obtained by summing
all relevant diagrams up to a certain order)  depends both on the color and the
spacetime quantum numbers  of the particles involved.
Suppose $\C_{v}$ is a complete set of independent color tensors, then
these two dependences can be separated by
a color decomposition
$$A_{tot}=\sum _{v}\C_{v} a_{v}\ .\eqno(3.1)$$
This serves to isolate  the colorless gauge-invariant subamplitudes $a_{v}$.

The amplitude $A$ for any diagram can be decomposed in a similar way. This
 is accomplished by decomposing the factors $S_{k}(q,p)$ in (2.1),
(2.4), and (2.14):
$$S_{k}=\sum _{v}\C_{v}s_{kv}\ .\eqno(3.2)$$
{}From Feynman rules, it is easy to see that when four-gluon vertices
are absent,  the color content of $S_{k}$ simply
factorizes. In other words, in that case all
$s_{kv}$ are proportional to some common $s_{k}$. In $U(N_{c})$
gauge theories the proportionality
constants are in fact either 0, 1, or $-1$; in $SU(N_{c})$ powers of
$N_{c}^{-1}$
also appear.  If four-gluon vertices are present, then there is a more
complicated dependence of $s_{kv}$ on $v$. This will be discussed in Sec.~4.3.

A convenient set of color tensors $\C_{v}$ will be discussed in Sec.~3.1 below.
These are the generalizations of the Chan-Paton factors [34] used
in the string theory. For a given diagram only some of these
color tensors are relevant;
a method to determine the relevant ones will be considered in
Sec.~3.2.

We shall also assume in the text  that we are dealing with
a $U(N_{c})$ gauge theory. The rules developed below are still valid for
$SU(N_{c})$
provided linkage of quark lines are always carried through
by more than one gluons.
Otherwise, $O(N_{c}^{-1})$ correction term are present. This and other
details will be discussed in Appendix B.

\bigskip\bigskip
\noindent {\bf 3.1 Color Tensors}

Let $T^{a} \ (a=0,1,\cdots,N_{c}^{2}-1)$ be the generators of $U(N_{c})$
in the fundamental (quark) representation,
 normalized so that
$$\Tr(T^{a}T^{b})
=\delta _{ab},\quad T^{0}={1\over\sqrt { N_{c}}}{\bf 1}\ .\eqno(3.3)$$
This normalization differs from the usual
one by a factor $\h$ on the rhs. It is convenient for certain purposes but
brings grief in others.
For $N_c=3$, these generators are
related to the Gell-Mann $SU(3)$ matrices by $T^a=\lambda^a/\r2$, which
means that the structure constant $f^{abc}$ used here, defined
by $[T^a,T^b]=if^{abc}T^c$, also differs from the usual one by a factor
$\r2$. The net result is that the QCD coupling constant $g$ used in
this article is $1/\r2$ of the usual coupling constant.

Let $\{a_{1},\cdots,a_{n}\}$
be the color indices for the $n$ (outgoing) gluons,
and $\{i_{1},\cdots,i_{m}\}$, $\{j_{1},\cdots,j_{m}\}$ be
respectively the color indices for the outgoing and incoming quarks.
A convenient set of color tensors $\C_{v}$ can be obtained as follows.

Let $v$ be any permutation of the set  of gluon indices $a_{i}$, written in the
cycle form:
$$v=(a'_{1}a'_{2}\cdots a'_{\l_{1}})\cdots(a'_{\l_{c-1}+1}\cdots a'_{\l_{c}})\
,\eqno(3.4)$$
where $\l_{i}$ are lengths of the $i$th cycle,
so that $\sum _{i}\l_{i}=n$, and $c$ is the
number of cycles in the permutation $v$.   For example, for $n=6$, one such
possibility might  be $v=
(a_{3}a_{5})(a_{2}a_{1}a_{6})(a_{4})$. The color tensors $\C_{v}$ in the
absence of fermions
can be taken to be
$$\C_{v}=
\Tr(a'_{1}a'_{2}\cdots a'_{\l_{1}})\cdots\Tr(a'_{\l_{c-1}+1}\cdots
a'_{\l_{c}})\
.\eqno(3.5)$$
In (3.5), a shorthand notation is adopted in which
a color index $a$   stands for the
generator $T^{a}$. Note that a cycle is
invariant under cyclic permutation,
and so is the trace `Tr', as it should be.

In the presence of fermions,  the construction of $\C_{v}$ is more
complicated. In this case they are of the form
$$\C_{v}=\prod _{k=1}^{m} (T^{b_{k}})_{i_{k}j'_{k}}C^{0}_{v'}\ ,\eqno(3.6)$$
where $\{j'_{1},\cdots,j'_{m}\}$ is a permutation of $\{j_{1},\cdots,j_{m}\}$,
and $\{b_{1},\cdots,b_{m}\}$ together with the color
indices in $v'$ form a permutation of
the gluon indices $\{a_{1},\cdots,a_{n}\}$. The symbol $C^{0}_{v}$ stands for
the
tensor $\C_{v}$ in (3.5).

\bigskip
\noindent {\bf 3.2 Feynman diagrams}

The relevant color tensors $\C_{v}$ appearing in a given Feynman diagram
can be read off from the color paths of the diagram [26]. A formal
definition of the color paths is given immediately below, but it is perhaps
far easier to look at the examples first before examining the
formal definitions.

A {\it color path} is a continuous path along the lines
of the diagram drawn in such a way that
\item{1.} every quark line is traversed {\it at most} once and
every gluon line as well as every ghost line  is traversed {\it at most}
twice;
\item{2.} paths along  quark lines are directed
along the arrow of  the quark lines. Paths along gluon lines can move
in either directions;
\item{3.} at a quark-gluon vertex junction, path along a quark line
turns to follow a gluon line and vice versa;
\item{4.} a path may be open or closed. If open, it must start
at an incoming quark and end at an outgoing quark. In particular, a path
cannot end at an external gluon line, so when it comes to the end of such a
line it must reverse and retrace its path.

A {\it color covering}
of the diagram is a collection of color paths so that every quark
line in the diagram is traversed {\it exactly} once,
and that every gluon line and
every ghost line in the diagram is traversed {\it exactly} twice.

\item{5.} for each color covering,
there is a predetermined order of paths through
 every three-gluon, four-gluon, or  ghost-gluon vertex junction, as
explained below.

At each three-gluon or ghost-gluon vertex, three lines with colors $a,b,c$
are involved. There are two ways to cyclically order these three colors:
$(abc)$ and $(acb)$. Each ordering fixes an order of the paths passing
through this vertex.
 For example, in the ordering $(abc)$, the path leading
into the vertex on line $a$(respectively $b,c)$
must exit the vertex on line $b$(
respectively $c,a)$.
Different orderings correspond to different paths and different
color coverings. Similarly, at a
four-gluon vertex, there are six ways to cyclically order the four colors
$a,b,c,d$ and consequently the corresponding paths through the vertex.
The ordering $(adcb)$
for example specifies that the path enetering the vertex along line $a$(
respectively $b,c,d)$
must leave along line $d$(respectively $a,b,c)$.

Let us look at some concrete examples. In Fig.~1,
 color paths $(p_{2}q_{2}q_{3}q_{4}p_{1})\equiv v_{1}$ and
 $(p_{4}q_{2}q_{1}q_{4}p_{3})\equiv v_{2}$
 together form a covering of the Feynman
diagram. In these notations, momenta read from right
to left are used to label  paths.

An example involving the presence of three- and four-gluon vertices
is shown in Fig.~2. In that case there are several color coverings,
one of which being $(p_{2}q_{7}q_{8}q_{5}q_{11}q_{9}q_{1}q_{10}$\break
$q_{12}p_{3})\equiv v_{1}$,
$(p_{4}p_{5}p_{5}q_{6}q_{8}p_{7}p_{7}q_{7}q_{2}q_{9}q_{10}p_{1})\equiv
v_{2}$, and $(q_{4}p_{6}p_{6}q_{3}q_{12}q_{11}) \equiv v_{3}$.
In this case, the ordering at the three gluon vertex is $(q_7q_8p_7)$
(momenta are used here to represent the corresponding colors) and
the ordering at the four-gluon vertex is $(q_{11}q_9q_{10}q_{12})$.

To each color path $v_{a}$ is associated a  color tensor
$\C_{v_{a}}$.
If $v_{a}$ is closed,
then $\C_{v_{a}}=\Tr(\cdots)$, where $\cdots$ stands for a
succession of the color indices of the {\it external gluons},
written from right to left in the order that the path encounters.
If $v_{a}$ is open, leading from
an incoming quark line with color index $j$ to
an outgoing quark line with color index $i$,  then $\C_{v_{a}}=(\cdots)_{ij}$.
If $\{v_{1},v_{2},\cdots,v_{u}\}$
is a path-covering of the Feynman diagram, then the total color factor
for the amplitude $A$ is
$\C_{v}=\prod _{a=1}^{u}\C_{v_{a}}$.
Notice that these tensors are of the form (3.6).

For example, for Fig.~1,
$$
\C_{v_{1}}=({\bf 1})_{ij}\ ,\quad \C_{v_{2}}=({\bf 1})_{i'j'}\ ,\eqno(3.7)$$
and for the color covering of Fig.~2 chosen before,
$$\C_{v_{1}}=({\bf 1})_{ij'}\ ,\quad \C_{v_{2}}=
(ac)_{i'j}\equiv (T^{a}T^{c})_{i'j}\ ,\quad \C_{v_{3}}=\Tr(b)\equiv \Tr(T^{b})\
.\eqno(3.8)$$

This association of color tensors with color paths is derived for $U(N_{c})$
colors, but it remains valid for $SU(N_c)$ provided pairs of quark lines,
when linked, are always carried out by
more than one gluon lines at a time.
For example,  (3.7) is not valid for $SU(N_c)$
because each of the two linkages of the two quark lines is carried out only
by one gluon line. On the other hand, (3.8) is valid for $SU(N_c)$
as well because both the linkages contain more than one gluon lines:
they are linked by two and four internal gluon lines respectively.

\bigskip\bigskip
\centerline{\bf 4. Spin Flow}

The factors $s_{kv}$ in (3.2) contains various scalar products and fermion
matrix elements like $\bar u\Gamma u$. Here $\Gamma $ is a scalar
function of the Dirac matrices $\gamma $, the external momenta
$p_{i}$, and the gluon/photon polarization vectors $\epsilon _{i}$, and $u,
\bar u$
are the fermion wave functions. From these expressions, one must reduce the
algebra to express the amplitude
$|A|$ in terms of measurable quantities, {\it i.e.,}
the scalar products of the external momenta and the helicities of external
particles. In the case of $S_{k}$ or $s_{kv}$,  Feynman
parameters will be involved as well. This task of algebraic reduction is not at
all easy for gauge theories
because of the rather complicated dependences of $u$ and $\epsilon $ on the
measurable
quantities, and because of the involvement of the non-commutative Dirac
algebra. Fortunately, for high energy processes where the masses of the
{\it external} particles can be ignored, such calculations can be immensely
simplified by using the {\it spinor helicity technique} [1--21].
The main purpose of
this section is to discuss how this technique can be applied to obtain $S_{k}$.
We shall assume in the following $d=4$. If $d\not=4$, the usual
complication of $\gamma _{5}$
and of the polarization vector [36]
arise and they have to be dealt with in the usual way.

The main idea of this technique is explained in Sec.~4.1, leaving the
detailed algebra and justification to Appendix C. Its application to
spinor QED and QCD is explained in Secs.~4.2 and 4.3, via a graphical technique
making use of {\it spinor paths}. Spinor paths are very similar to color
paths owing to the resemblance between the color algebra and the Dirac
algebra. Yet in detail the two kinds of  paths are different because of the
different dependences of the Feynman rules on color and  spin.

\bigskip
\noindent {\bf 4.1 Spinor Helicity Technique}

The involvement  of  Dirac algebra can be traced back to
a mixing of helicity and chiral eigenstates. An external (massive)
fermion carries a definite helicity, but not a definite chirality.
On the other hand, it is chirality that is conserved at QED
and QCD vertices. This forces a mixing of helicity states
whenever interactions take place, which brings about a multichannel
problem described by the  $4\times 4$
Dirac matrices. However, if the external fermion is massless, or can be
regarded as such
in a high energy process, then helicity and chirality are essentially
the same,  no mixing of states need to occur,  the four-channel
problem
is reduced to a one-channel problem and the complicated Dirac algebra
can thus be avoided. This is the basic idea of the spinor
helcity technique [1--21].

This technique is also useful for spin-1 particles, because
kinematically a spin-1 particle may be thought of as a composite
 of two spin-${1\over 2}$
particles.
Moreover, the polarization vector of a gauge particle is gauge dependent, so
in principle it is possible to simplify calculations by a suitable
gauge choice.
This turns out to be so in the spinor helicity technique.

Unless otherwise specified, it
will be assumed from now on that all
external momenta $p_{i}$ are massless. The massless
spin-${1\over 2}$ wave functions
$u_{\lambda }(p_{i})$ and $\bar u_{\lambda }(p_{i})$  will be written
respectively as $\ket{p_{i}\lambda }$ and $\bra{p_{i}\lambda }$, where $\lambda
=\pm
1$ is (twice) the helcity of the fermion. Using the fact that these are also
$\gamma _{5}$ eigenstates, one concludes immediately that
$$\bk{p_{i}\pm |p_{j}\pm }=0\ .\eqno(4.1)$$
The non-zero matrix elements are conventionally written as
$$\bk{p_{i}p_{j}}\equiv \bk{p_{i}-|p_{j}+},\quad \sbk{p_{i}p_{j}}\equiv
\bk{p_{i}+|p_{j}-}\
,\eqno(4.2)$$
whose normalization and  charge conjugation properties  can be
chosen so that
$$\eqalignno{
\bk{p_{i}p_{j}}&=-\bk{p_{j}p_{i}}\ ,&(4.3)\cr
\sbk{p_{i}p_{j}}&=-\sbk{p_{j}p_{i}}\ ,&(4.4)\cr
\bk{p_{i}p_{j}}\sbk{p_{j}p_{i}}&=2p_{i}\.p_{j}\ ,&(4.5)\cr
\sbk{p_{j}p_{i}}&={\rm sign}(p_{i}\.p_{j})\bk{p_{i}p_{j}}^{*}\ .&(4.6)\cr}$$
See Appendix C for the derivation of these
and other related formulas.

The polarization vectors for an outgoing photon/gluon of momentum $p$
and helicity $\pm 1$ can be chosen to be
$$\epsilon _{\pm }^{\mu }(p,k)=\pm {\langle p\pm |\gamma ^{\mu }|k\pm \rangle
\over
\sqrt {2}\langle k\mp |p\pm \rangle }\ ,\eqno(4.7)$$
where $k$ is an arbitrary massless momentum known as the {\it reference
momentum}. Polarization vectors with different $k$'s are related by
gauge transformations (see (C21)).
There are two advantages with this form of the
polarization vector: it is expressed in terms of spinor notations so
the spinor helicity technique can be used; it is also symmetrical
in $p$ and $k$ so that $k\.\epsilon _\pm(p,k)=0$ as well as
$p\.\epsilon _\pm(p,k)=0$, making it possible to simplify the
calculation by a proper choice of $k$.

It turns out that  all matrix elements of the Dirac matrices and all
expressions involving the polarization vectors can be reduced to the
bracket quantities in (4.2), which according to (4.3)--(4.6),
are essentially the square roots of the dot momenta. In this way the
objective of expressing everything in terms of measurable quantities
is realized.

Graphical methods can be designed to
read off $S_k$ directly from the Feynman diagram. This  will
be illustrated in the next two sections for massless QED and QCD,
but other rules can be formulated to cover more complicated situations [26].
Feynman gauge  will be used throughout this paper for gauge
propagators .

\bigskip
\noindent {\bf 4.2 Spinor QED}

In order to describe how $S_{0}(q,p)$ is computed graphically, it is necessary
to
discuss what a {\it spinor path} and a {\it spin factor} are. Again,
it might be easier to follow the illustrative examples   before
returning to the formal definitions immediately below.

Consider first diagrams with no external photons present.
A {\it spinor path} is a continuous path along the lines
of the diagram drawn in such a way that
\item{1.} every electron line is traversed {\it at most} once and
every photon line as well as every ghost line  is traversed {\it at most}
twice;
\item{2.} a path along an electron line must travel in the same (opposite)
direction of the previous electron line the path
 went through if these two electron
lines carry opposite (the same) helicities.
Paths along internal photon lines can move
in either directions;
\item{3.} a path along an electron line must turn to follow an internal photon
line, and vice versa, when an electron-photon  vertex is encountered;
\item{4.} a path may be open or closed. If open, it must start
and end at external electron lines.

In other words, in the absence of photons, a spinor path in QED is almost
identical
to a color path in QCD, except for the fact that the spinor path can have both
orientations along an electron line depending on the helicities involved.
We will see that more differences will emerge when external photons are
present, and when other vertices are encountered in the case of QCD.

A {\it spinor covering} of the diagram is a set of spinor paths so that
every electron line in the
diagram is traversed {\it exactly once}
 and every photon line {\it exactly twice} by these paths.

For example, Fig.~3
 is covered by three spinor paths,
$(p_{2}q_{5}q_{1}q_{7}p_{3})\equiv A$, $(p_{4}q_{6}q_{3}q_{5}p_{1})\equiv B$,
and
$(q_{4}q_{6}q_{2}q_{7})\equiv C$, in which the paths are labelled by the
momenta
of the lines they pass through, to be read from right to left. Paths
A and B are open, and path C is closed. It is easy to see from the above
list that every electron line $(p_{i},q_{i},1\le i\le 4)$ is traversed exactly
once
and every internal photon line ($q_{5},q_{6},q_{7}$) is traversed exactly
twice.
Note that if the upper electron line carried negative helicity, then
the diagram will be covered by only two paths $A'$ and $B'$:
$(p_{4}q_{6}q_{2}q_{7}p_{3})\equiv A'$ and
$(p_{2}q_{5}q_{1}q_{7}q_{4}q_{6}q_{3}q_{5}p_{1})\equiv B'$.

To each spinor path A is associated a {\it spin factor} ${\cal G}_{A}$
obtained by pairing consecutive momenta along the path, alternately with
brackets $\bk{\cdots}$ and $\sbk{\cdots}$. In this pairing, the momentum
$q_{r}$
of an {\it internal
fermion}  always occurs {\it twice}, as in $[\cdot q_{r}]\bk{ q_{r}\cdot}$,
or in $\bk{\cdot q_{r}} [q_{r}\cdot]$, but the momentum of any
 {\it internal gluon} line is to be {\it omitted}.
For an open path, one can start the pairing process either from the
end of the outgoing electron, or from the end of the incoming electron.
In the former case, the pairing proceeds from left to right, starting
with a $\langle \cdots$ ($[\cdots$) when the outgoing electron has a
negative (positive) helicity. In the latter case, the pairing proceeds
from right to left, starting with a $\cdots\rangle $ ($\cdots ]$) when the
incoming electron has a positive (negative) helicity. If the path
is closed, then one can start at any point along the path, but
both helicities should be used and a sum taken.

For example, the spin factors for Fig.~3 are
$$\eqalignno{
 {\cal G}_{A}&=\sbk{p_{2}q_{1}}\bk{q_{1}p_{3}}\ ,&\cr
{\cal G}_{B}&=\sbk{p_{4}q_{3}}\bk{q_{3}p_{1}}\ ,&\cr
{\cal G}_{C}&=\sbk{q_{4}q_{2}}\bk{q_{2}q_{4}}
+\bk{q_{4}q_{2}}\sbk{q_{2}q_{4}}=4q_{2}\.q_{4}\ .&(4.8)}$$
If instead the upper electron has negative helicity, then
$$\eqalignno{
{\cal G}_{A'}&=\bk{p_{4}q_{2}}\sbk{q_{2}p_{3}}\ ,&\cr
{\cal G}_{B'}&=\sbk{p_{2}q_{1}}\bk{q_{1}q_{4}}\sbk{q_{4}q_{3}}\bk{q_{3}p_{1}}\
.&(4.9)}$$

The momentum pairs appearing in (4.2) are massless, but in the pairings above
off-shell momenta $q_{r}$ appear. In their presence the brackets
are defined  recursively as follows:
$$\eqalignno{
\bk{p_{i}q_{r}}\sbk{q_{r}p_{j}}&\equiv \sum _{k}
I_{rk}\bk{p_{i}p_{k}}\sbk{p_{k}p_{j}}\ ,&\cr
\sbk{p_{i}q_{r}}\bk{q_{r}p_{j}}&\equiv \sum _{k}
I_{rk}\sbk{p_{i}p_{k}}\bk{p_{k}p_{j}}\ ,&(4.10)}$$
where $I_{rk}$ are the coefficients appearing in (2.3).
Note that only bilinear combinations as appearing on the left hand side
of (4.10) are defined, and this will always be the case for the brackets
appearing in the spin factors ${\cal G}_{A}$.
Single unpaired $\bk{qq'}$ or $\sbk{qq'}$ for off-shell $q,q'$
remain undefined. Note also that if $q_r=\kappa p_k$ for some
constant $\kappa $ and some massless momentum $p_k$, then
(4.10) is perfectly consistent with eq.~(C4).

So far we have considered diagrams without external photons. In their
presence, spinor paths and spin factors can be calculated by
modifying the diagram in the following way. Attach a short
fermion line to every external photon line and assign the following
characteristics to these artificial fermions. The helicity of the artificial
fermion is equal to the helicity of the photon, the momentum of the
incoming artificial fermion is equal to the reference momentum of the photon,
and the momentum of the outgoing artificial fermion is equal to the actual
momentum of the photon. For example, the modified diagram for Fig.~4 is given
in Fig.~5.

The modified diagrams should now be considered as a Feynman diagram without
external photon lines, and their  spinor paths as well as their spin
factors ${\cal G}_{A}$ should be calculated just as before. For example, the
spin
factors for Figs.~4 or 5 are
$$\eqalignno{
{\cal G}_{A}&=\sbk{p_{2}q_{1}}\bk{q_{1}p_{3}}\ ,&\cr
{\cal G}_{B}&=\sbk{p_{4}p_{5}}\ ,&\cr
{\cal G}_{C}&=\bk{k_{5}q_{9}}\sbk{q_{9}q_{3}}\bk{q_{3}p_{1}}\ ,&\cr
{\cal G}_{D}&=\bk{p_{6}q_{6}}\sbk{q_{6}q_{5}}\bk{q_{5}q_{8}}\sbk{q_{8}k_{6}}\
,&(4.11)}$$
where $k_{5},k_{6}$ are respectively the reference momenta for the photons with
momenta $p_{5}$ and $p_{6}$. In actual calculations, $k_{5},k_{6}$ are usually
chosen
to be one of the $p_{i}$'s.

In terms of these spin factors, $S_{0}(q,p)$ in (2.4) is given by
$$S_{0}(q,p)={\cal N}{\cal G}\ ,\eqno(4.12)$$
where ${\cal G}=\prod _{A}{\cal G}_{A}$ is  the product of the spin
factors ${\cal G}_{A}$ over all the paths $A$ in a spinor covering.
If the diagram should
have more than one spinor coverings (which happens in QCD but never in QED),
then a sum over all the coverings should be taken as well.

The factor ${\cal N}$ for a diagram
with a symmetry factor $ S_{sym}$, $\l_F$
fermion loops, $n_{\gamma }$ external and $N_{\gamma }$  internal photon lines
is
\b{
{\cal N}&=(\r2 e)^{n_{\gamma}+2N_{\gamma}}(-1)^{\l_F}S_{sym}^{-1}
\prod _{a=1}^{n_{\gamma }} {\cal F}_{a}\ ,&(4.13)\cr}$$
where
$$
{\cal F}_{a}= \cases{+1/\bk{k_{a}p_{a}}, &($\lambda _{a}=+1$)\ ;\cr
-1/\sbk{k_{a}p_{a}},&($\lambda _{a}=-1$)\ ,\cr}\eqno(4.14)$$
with $p_{a},k_{a},\lambda _{a}$ being respectively the actual momentum,
the reference momentum, and the helicity of the $a$th external photon.

Similar graphical rule can be developed to calculate $S_{k}$. This involves
shrinking the contracted lines  and making
new links according to (2.6).

\bigskip
\noindent {\bf 4.3 QCD}

The QCD colorless factors $s_{0v}$ in (2.14) is like the factor
$S_{0}$ in QED, if we make the obvious substitution of quarks and gluons
for electrons and photons. It is given
by the right-hand side of (4.12), with ${\cal N}$ and
${\cal F}_a$ given by
 (4.13) and (4.14). The factor ${\cal G}$ is now dependent on the color
path $v$ if four-gluon vertices are present, and it is equal to
$${\cal G}=\sum \prod _{A}{\cal G}_{Av}\ ,\eqno(4.15)$$
where the sum is taken over all the spinor coverings of the diagram.
The spinor factor ${\cal G}_{Av}$ depends on the color path $v$ only when
multi-gluon or ghost-gluon vertices are present. Otherwise, ${\cal G}_{Av}$
is calculated exactly like QED.

Before discussing how spinor paths are constructed in the presence of these
non-abelian vertices, let us first digress to discuss
{\it (color-)oriented vertices} [26] in the  {\it
background
Feynman gauge} [37]. These are shown in Fig.~6.

A {\it background gauge} is one in which the gauge-fixing
depends explicitly on
a background Yang-Mills field from which external gluon lines are extracted.
As a result,
vertices containing external gluon lines
(indicated by a circled `A' in Fig.~6) have different Feynman
rules than vertices without. Furthermore, extra ghost-gluon vertices appear,
as shown in Fig.~6, where
a thin solid line represents a gluon ($g$), a thick solid line
a quark ($q$), and a dashed line a ghost ($G$).  A background
{\it Feynman gauge} is one in which Feynman gauge is used in the gluon
propagators.

An {\it oriented vertex} differs from an ordinary vertex in that the
lines in the vertex diagram are ordered in a clockwise order, and
the vertex Feynman rules depend on the line orderings. It is the appropriate
vertex to use when color factors are taken to be the Chan-Paton
tensors of Sec.~3.1 rather than the usual ones.  Their Feynman rules
can be derived from the ordinary Feynman rules by using eqs.~(B3)
and (B4).  In particular, those in the background Feynman gauge
can be derived this way  from Fig.~1 of the first paper in [37]
by setting $\alpha=1$.

The Feynman rules for Fig.~6 are indicated below the diagrams, and they
should be interpreted in a particular way.
A line with a heavy dot at the end is a {\it dislodged line}.
The Feynman-rule factors for diagrams with dislodged vertices
are always vectors, whose component is indicated by the Lorentz index
of the dislodged line. A continuous gluon line signifies a $g_{..}$ factor.
In Fig.~6 the color and coupling-constant factors
are understood: there should always be a coupling constant $g$
for every cubic diagram and a $g^2$ for every quartic diagram.
The color factor for Fig.~6(t) is $T^a$, and that of any
other diagram is $\Tr[a_1a_2a_3(a_4)]$, where $a_1,a_2,a_3,(a_4)$
are the color indices of the $g$ or $G$ lines read in clockwise order.
Finally, as mentioned before,
an {\it external} gluon line  is indicated
by a circled `A'. The gluon lines without a circled `A' can be
either an external or an internal gluon line, unless such a configuration
is already explicitly represented by a diagram in Fig.~6 with circled `A's.

To give some examples, let
the (outgoing momentum, Lorentz index, color index) of line 1
be $(q_1,\alpha,a)$, that of line 2  be $(q_2,\beta,b)$,
etc. An external momentum will be designated as $p_i$ rather than $q_i$.
Fig.~6(b) is a vertex for one
external and two internal gluon lines, with factor
$g\Tr(abc)(q_{2}-q_{3})_{\alpha }
g_{\beta \gamma }$. Similarly,
Fig.~6(e) is a vertex for four internal gluon lines, with vertex factor
$g^2\Tr(abcd)(+2)g_{\alpha \gamma }g_{\beta \delta }$.

Note that all topologically different permutations must be taken into
account. For example, in the usual four-gluon vertex, there
are six terms with a color factor like $f^{abe}f^{cde}$ in each.
Using eq.~(B4), this generates 24 terms. All these 24 terms are summarized
in Figs.~6(e) and 6(f) as follows. There are six possible permutation
of color indices in each diagram. For a fixed permutation, there
are two different diagrams of type 6(f), {\it e.g.,}
$\Tr(abcd)g_{\alpha\beta}g_{\gamma\delta}$ and
$\Tr(abcd)g_{\delta\alpha}g_{\beta\gamma}$. The factor of $+2$ is
6(e) also represents two terms. Altogether there are thus
$(2+2)\times 6=24$ terms, as it should be.

The main advantage of the background-gauge is that in Figs.~6(c) and 6(d),
where an internal line is dislodged, the vertex factor depends only on
the {\it external} momentum. This simplifies matters  in loop calculations
because otherwise internal momenta are involved,  they
depend on the Feynman (Schwinger) parameters
and must be calculated using (2.9), and they are also subject to contractions
(2.6). For low order calculations with many external photons/gluons,
this simplification far outweighs the complications brought in by
the presence of the extra vertices in Fig.~6.

If ordinary
(not background) gauges are used, then
there should be no distinction between  internal and
external gluon lines. All diagrams containing a circled `A' in Fig.~6 can be
put to zero. In particular, there is only one triple-gluon vertex, that
of Fig.~6(a).

{}From Fig.~6 it is possible to deduce how
spinor paths and spinor coverings are constructed in the presence of the extra
QCD gluon vertices.
First modify
the $(ggg)$ vertices in a Feynman diagram by dislodging one of the three
gluon lines. There are obviously three different ways of doing
so and these correspond to different spinor paths. Next,
pair up the four gluon lines in a $(gggg)$ vertex into two pairs,
 similar to what is shown in
Figs.~9(e)--(h). There are again three ways of doing so and
these again correspond to different spinor paths.

The rules for constructing a spinor path are identical to the rules
in QED, with the obvious replacement of electrons by quarks and photons by
gluons. The following addition rule tells us what to do when a non-abelian
vertex is encountered:
\item{5.} a dislodged gluon line should be regarded as if it were an external
gluon line;  a paired gluon line should be treated as a single continuous line.

The spinor factor ${\cal G}_{Av}$ in (4.15) is determined in exactly the same
way as in the QED case, but with the following addendum. The artificial fermion
line attached to a dislodged gluon line should carry initial and final momenta
equal to the vector given below the appropriate diagram
in Fig.~6. The helicity of
the artificial fermion is immaterial and can be chosen to be either.
Additional numerical factors appearing in Fig.~6 and the
appropriate number of coupling constants should also be incorporated into
${\cal G}_{Av}$.
Note that there is very little coupling between the color path $v$ and the
spinor path $A$: only a sign and perhaps a numerical factor.

Now examples.
There is only one color covering each for Figs.~1
and 3,
and the ${\cal G}$ factors were given in (4.8) and (4.11) respectively.

Consider Fig.~2 with the color path discussed in Sec.~3.2. Fig.~7 indicates
a spinor covering for a possible helicity configuration,
where $k_{i}$ are the reference momenta of the external gluons with actual
momenta $p_{i}$. The spinor factors in the background gauge are
$$\eqalignno{
{\cal G}_{A}&=\sbk{p_{2}k_{7}}\ ,&\cr
{\cal G}_{B}&=\bk{p_{7}q_{2}}\sbk{q_{2}q_{5}}\bk{q_{5}(2p_{7})}\ ,&\cr
{\cal G}_{C}&=\sbk{(2p_{7})q_{6}}\bk{q_{6}k_{5}}\ ,&\cr
{\cal G}_{D}&=\sbk{p_{5}p_{4}}\ ,&\cr
{\cal G}_{E}&=\bk{p_{3}p_{1}}\ ,&\cr
{\cal
G}_{F}&=(-1)\bk{p_{6}q_{3}}\sbk{q_{3}q_{1}}\bk{q_{1}q_{4}}\sbk{q_{4}k_{6}}\
.&(4.16)}$$

\bigskip

\centerline{\bf 5. String-like Formulas}

A one-loop $n$-photon/gluon amplitude can be computed using superstring
and first quantized techniques [22--25]. The result resembles (2.14),
but the numerator function $S(q,p)$ differs
from what one would get by  using (2.6)--(2.11) directly. The purpose of this
section is to show that these {\it stringlike formulas} can also be
obtained by purely field-theoretic means [27], and in this way they can
 be generalized to
multiloop amplitudes and to all processes.

Let $a$ be a cubic vertex in scalar QED, with an external photon line
as shown in Fig.~8. Its vertex factor is
$$C_a=e \epsilon _a\.(q_{a'}+q_{a''})\ ,\eqno(5.1)$$
where $\epsilon _a$ is the polarization vector of the outgoing photon.
What the stringlike reorganization achieves, see Sec.~5.1,
is to replace these factors and their contractions
by derivatives wrt the Schwinger parameters.  The advantage
of this new form, compared to the direct use of (2.9)--(2.11), is that
different terms are connected by derivatives so that
integration by parts [22--25] can be used to relate them.

For spinor QED, the vertex factor is changed to
$$F_a=e\epsilon _a\.\gamma \ ,\eqno(5.2)$$
and there are additional momentum dependence from the spinor propagators.
By using the Gordon identity, $F_a$ can be partially converted to $C_a$
so the stringlike formulas for scalar QED can be applied.
See Sec.~5.2. However, Gordon identity changes chirality-conserving
amplitudes to chirality-violating amplitudes, so this
change generally does not mesh well with the spinor helicity technique.

For QCD, a three-gluon vertex has a term   resembling (5.1), but
it also has two other terms. In the background gauge discussed in
Sec.~4.3, the other two terms  only depend on the external momenta,
so the stringlike reorganization for scalar QED is once again applicable
here. See Sec.~5.3.

{}From a purely field-theoretical point of view, these new formulas
  can be motivated by, and derived from,
the desire to restore manifest local gauge invariance. To do so
considerable   knowledge of the electric circuit theory is required.
These details will be supplied in Appendices
A and D.

For topological reasons, these stringlike reorganizations can be
carried out only at external vertices. Whether other
simplifications exist for internal vertices
is presently unknown.

\bigskip
\noindent{\bf 5.1 Scalar QED}

The scalar particles
considered here are allowed to interact `strongly' among themselves in
any non-derivative manner.
Their interactions with the photon field $A_\mu $
are given by the cubic vertex $C^\mu A_\mu $ of Fig.~8 and the
quartic (seagull) vertex $Q^{\mu \nu} A_\mu A_\nu $ of Fig.~9, where
\b{
C^\mu &=e (q''+q')^{\mu }\ ,&(5.3)\cr
Q^{\mu \nu }&=e^{2} g^{\mu \nu }\ .&(5.4)}$$
The numerator of photon propagator in covariant gauges is
$$-g^{\mu \nu }+\xi p^{\mu }p^{\nu }/p^{2}\ .\eqno(5.5)$$
Unless otherwise specified, the Feynman gauge $\xi =0$ will be used
throughout.

Consider a QED diagram with $n_e$ external vertices, of which $n_3$
are cubic vertices of the type (5.1), with outgoing photon momenta
$p_a$ and polarization $\epsilon _a$. Let $S_0^{ext}(q,p)$
be the product of these $n_3$ vertex factors, and the numerator function
be $S_0(q,p)=S_{0}^{ext}(q,p)S_{0}^{int}(q,p)$.
Then the scattering amplitude is given by (2.14) to be
\b{A&=\int [D_S\alpha ]
\Delta ^{-d/2}S(q,p)\exp[-i(M- P)]\ ,&(5.6)\cr}$$
which turns out to equal to
\b{
A&=\int [D_S\alpha ]
\Delta ^{-d/2}\{S^{int}(\tilde q,p)\exp[-i(M- \tilde P)]\}_{ML}\ ,
&(5.7)}$$
where $S^{int}=\sum_kS^{int}_k$ is the total numerator factor
with the $n_3$ vertices removed.
If $q_r$ and $P$ are given by (2.3) and (2.5), $a,b$ are external
vertices like Fig.~8, then
\b{\tilde P&=\sum_{a,b=1}^{n_e}(p_a
+ie \eta_a\epsilon _a\p_a)\. (p_b+ie\eta_b\epsilon _b\p_b)
Z_{ab}(\alpha )\ ,&(5.8)\cr
\tilde q_r&=\sum_{a=1}^{n_e}(p_a+ie\eta_{ar}\epsilon _a\p_a)
I_{ra}(\alpha )\ ,\qquad {\rm with}&(5.9)\cr
\p_a&\equiv {\p\over\p \alpha _{a'}}-{\p\over\p \alpha _{a''}}\ .&(5.10)
}$$
 $\eta_a$
is defined to be 1 for one of these $n_3$ vertices and 0 otherwise.
Similarly,
$$\eta_{ar}=\cases{1, &if $\eta_a=1$ and line $r$ is not adjacent to vertex
$a$;
\cr
2, &if $\eta_a=1$ and $a$ is at one end of the line $r$;\cr
0, & if $\eta_a=0$.}\eqno(5.11)$$
The subscript $ML$ in (5.7) stands for {\it M}ulti{\it L}inear. It means
that only the terms multilinear in
the $n_3$ polarization vectors
$\epsilon _a$ should be retained when the curly bracket is expanded.

Eq.~(5.7) is the result of the {\it stringlike reorganization}. It
is valid provided the impedance matrix $Z_{ab}$ is given in the
{\it diagonal scheme} of (A16), where $Z_{aa}=0$ for every
$a$. In this  scheme (see Appendix A
for more details), the impedance matrix can be computed using the
graph-theoretical formula [see (A18)]
$$Z_{ab}=-\h\sum _{T_{2}^{ab}}\prod ^{\l+1}\alpha \ ,
\quad(a\not=b)\ ,\eqno(5.12)$$
where $T_{2}^{ab}$ is the set of all 2-trees with vertices $a$ and
$b$ belonging to two separate 1-trees.

Sting-like organization
replaces all external cubic electromagnetic vertices by momentum
shifting as in (5.8) and (5.9). The $\alpha $-dependence of the
original and the shifted terms differ only by derivatives
$\p_a$ and $\p_b$, so they can be converted into one another by
an integration by parts. This can lead to simplification in calculations
[22--25].
In the special case of a one-loop
$n$-photon amplitude, this formula was first obtained from string
calculations. The formula (5.7) however is valid for multiloop amplitudes,
and for all scattering processes.

Eq.~(5.7) can be derived from an attempt to understand local gauge
invariance. Its derivation and the question of manifest gauge invariance
will be taken up in Appendix D.

\bigskip
\noindent{\bf 5.2 Spinor QED}

The vertex in spinor QED is given by (5.2). The main idea in stringlike
reorganization is to use the Gordon identity to convert it partially to
the scalar vertex (5.1), so that the
formulas in Sec~5.1 can be used once again. The Gordon identity on mass
shell is well known to be
$$\bar u_{\lambda '}(p')\gamma ^{\mu }u_{\lambda }(p)=
{1\over 2m}\bar u_{\lambda '}(p')[(p'+p)^{\mu }+i \sigma ^{\mu \nu}
(p'-p)_{\nu }]u_{\lambda }(p)\ .\eqno(5.13)$$
A similar off-shell formula exists [27]
and will be explained in Appendix E.
This formula allows the electromagnetic vertex $F^\mu A_{\mu }=e
\gamma ^{\mu }A_{\mu }$  to be replaced by a sum of
the scalar QED vertices, the cubic vertex
  $C^\mu A_{\mu }=e(q^{'\mu} +q^{''\mu} )A_\mu$,
the seagull vertex $Q^{\mu \nu} A_\mu A_{\nu }=e^2g^{\mu \nu}A_\mu A_\nu$,
and the magnetic-moment vertex
$$S^\mu A_\mu =-ieA_{\mu }\sigma ^{\mu \nu }p_{\nu}\ ,\eqno(5.14)$$
provided the spinor propagators $(m-\gamma q)^{-1}$ are simultaneously replaced
by the
scalar propagators $(m^2-q^2)^{-1}$. In addition, a factor $(2m)^{-1}$ should
be added to every pair of external fermion lines, and a factor $\h$
to every fermion loop.

In the original form, only helicity-conserving matrix elements survive
in the $m\to 0$ limit. See Sec.~4.2. In the modified form, propagators no
longer
contribute to $S_0(q,p)$, and the new vertices $C,Q,S$ all contain
 an even number of Dirac $\gamma $-matrices. Hence helicity-conserving
matrix elements tend to zero in the
$m\to 0$ limit, which is why the extra factors $(2m)^{-1}$ are present to
render the results finite, and
  to convert them into chirality-violating matrix elements.
See (C19) for exact formulas. This conversion
unfortunately goes against the grain
of the spinor helicity technique, which is to make maximum use of
chirality conservation. Nevertheless, when external fermion lines are
absent, chirality-conservation is not maximally used and the result
of the conversion remains to be simple.

Other than the magnetic-moment vertex $S^\mu $, which
by itself is gauge invariant,
everything else is identical
to scalar QED. Hence the scattering amplitude is similar to (5.7)
and can be written as
\b{
A&=\int [D_S\alpha]  \Delta ^{-d/2}\{ S^{int}(\tilde q, p)
\exp[-i(M- \tilde
P)]\}_{ML}\ ,&(5.15)}$$
where $ S^{int}_0(q,p)$ receives its contribution from
all the modified vertices  with the extra factors $(2m)^{-1}, \h$, etc.,
but with the $n_3$ external cubic vertices (5.1) removed.
As usual, $ S^{int}$
is the sum of $ S_0^{int}$ and its contractions. In the special
case when internal photon lines are absent ({\it i.e.,} one-loop
$n$-photon amplitudes), the magnetic-moment
vertices $S^\mu $ have no internal-momentum dependence, then
\b{
S^{int}&= S_0^{int}=\h(-ie)^{m}\Tr[(\epsilon _{a_1}\. \sigma \.
p_{a_1})(\epsilon _{a_2}\. \sigma \. p_{a_2})\cdots(\epsilon _{a_m}\.
\sigma \. p_{a_m})]&\cr
&\equiv\h\exp[-ie\sum_{i,j=1}^m \epsilon _{a_i}\.p_{a_j}
G_F^{a_ia_j}]_{ML} \ ,&(5.16)}$$
where $a_1,a_2,\cdots,a_m$ are the magnetic-moment vertices around
the fermion loop. Using (5.15) and (5.16), the final result in this
special case can be written as
\b{
A&=\h\int [D_S\alpha]  \Delta ^{-d/2}
\exp[-i(M- \tilde P)-ie\sum_{a,b=1}^{n_3} \epsilon _{a}\.p_{b}
G_F^{ab}]_{ML}\ ,&(5.17)}$$
which agrees with the results from superstring and first-quantized
calculations [22--25].

\bigskip
\noindent{\bf 5.3 Pure QCD}

There are three terms in a three-gluon vertex. In ordinary covariant
gauges, each of these three terms resembles (5.3). In the background
gauges discussed in Sec.~4.3, the same Feynman rule holds for internal
vertices but not the external ones. In the latter case, one of
the three terms, the one with the external gluon dislodged,
resembles (5.3), but the other two terms where an internal gluon line
is dislodged contain only the  momentum of the external
gluon. This means that
no contraction is needed for these other two terms, and the
stringlike reorgainizational techniques of scalar QED can once
again be used as follows.

It is helpful to use  spin paths to organize the many terms present.
Since quarks are absent in pure QCD,
there is no reason to use {\it spinor} paths; the simpler
 {\it vector paths}  would be sufficient.
A {\it vector path} is a continuous path along the lines
of the diagram drawn in such a way that
\item{1.} every gluon line is traversed {\it at most once};
\item{2.} a path may be open or closed. If open, it must start
and end at external gluon lines;
\item{3.} for this purpose a dislodged gluon line is treated like
an external line.

A {\it vector covering} is a collection of vector paths so that each
external and internal gluon line is traversed exactly {\it once}.

{\it Spin factors} are computed as follows [26].
To the end of each path a vector is assigned. For an external gluon
this is its polarization vector; for a dislodged gluon it is the vector
shown under the diagrams in Fig.~6.
The spin factor ${\cal G}_{A}$ for a vector path A is the dot product of
the vectors at the two ends of the path, if the path
 is open, together with whatever other numerical
factors present at the vertices along the path. If the path is
closed, then ${\cal G}_A=g^{\mu }_{\mu }=d$, times other factors in
Fig.~6 that
the path encountered. The total spin factor ${\cal G}$ is once again given
by the product of ${\cal G}_{A}$'s in the vector covering.

Except for a factor $(-e)$, the coupling in Fig.~6(b),
where the external gluon is dislodged, is identical to the
coupling (5.3) in scalar QED. For an external vertex $a$
in which the external gluon line is dislodged, the formulas
in Sec.~5.1 apply, {\it viz.,} $p_a$ is replaced by $p_a+ie\eta_{a(r)}
\epsilon _a\p_a$.

There are   vector paths where none of the external
gluons are dislodged. Some of them, such as those
in Figs.~10 and 11, may form a loop passing through $m$ external triple-gluon
vertices.
We will call these paths {\it $m$-loops}.
For an $m$-loop, vertices in Figs.~6(c) and 6(d) are relevant.
In particular, for something like Fig.~10, the spin factor is
\b{
{\cal G}&=\prod _{i=1}^{m}(-2e
\epsilon _{a_{i}}\.p_{a_{i+1}})&\cr
&=\exp\[-2 e\sum _{i=1}^{m}
\epsilon _{a_{i}}\.p_{a_{i+1}}\]_{ML}\ ,&(5.18)}$$
where $(a_{m+1}=a_{1},a_{2},a_{3},\cdots,a_{m})$ are the ordered external
vertices
around the $m$-loop.

If the other internal line at one or several
 vertices $a$ is dislodged instead,  Fig.~10 turns
into diagrams like Fig.~11. The result in (5.18) is to interchange
$\epsilon _a$ with $p_a$, with a minus sign. The sum of all such
contributions is
$${\cal G}_m=\exp\[2 e\sum _{i=1}^{m}(\epsilon _{a_{i}}-p_{a_i})\.
(\epsilon _{a_{i+1}}-p_{a_{i+1}})\]_{ML}\ .\eqno(5.19)$$

For a one-loop $n$-gluon amplitude made up of triple gluon vertices,
every external gluon line is either dislodged, or is a member of
a $m$-loop for some $m$. Thus its amplitude is given by
\b{
A=\int [D_S \alpha ]\sum _{m\le n}\exp[ &-iM+i\sum _{a,b=1}^n
(p_{a}+ie \epsilon _{a}\p_{a})\.
(p_{b}+ie \epsilon _{b}\p_{b})Z_{ab}(\alpha )&\cr
&+2e\sum _{i=1}^{m}
(\epsilon _{a_{i}}-p_{a_i})\.(\epsilon _{a_{i+1}}-p_{a_{i+1}})]_{ML}\ ,&(5.20)
}$$
where the sum in front of the exponential
 is over all $m$-loops with $m\le n$, with $(a_1,\cdots, a_m)$
arranged in the same cyclic order as the original vertices. This
formula agrees with the one obtained from string calculations.

The topology for a general Feynman diagram could be very complicated,
in which case
one has to resort back to the general formula (5.15), where
$S^{int}(q,p)$ now includes all the spin factors
 where the external gluon lines are not dislodged.

\bigskip \bigskip
\centerline{\bf Acknowledgement}

This work is supported in part by the Natural Science and Engineering
Research Council of Canada and the Qu\'ebec Department of Education.

\vfill\eject

\centerline{\bf Appendix A. Circuit theory}

Consider an electric circuit with $n$ vertices, each hooked up to an
outgoing current $p_{i}\ (1\le i\le n)$, and $N$ internal
lines, each carrying a resistance $\alpha _{r}\ (1\le r\le N)$ and a current
$q_{r}$.
If the line $r$ points from vertex $j$ to vertex $i$, then we write
$r=(ij)$ and say that the line $r$ and the vertices $i$ and $j$
are mutually {\it adjacent}.

Define a $(n \times N)$-dimensional {\it incidence matrix}, whose
elements
$A_{kr}$ are $+1,-1$ respectively if
$k=i,j$, and  is 0 if $k$ is neither $i$ nor $j$. Current
conservation at vertex $i$ is then given by the equation
$$p_{i}=\sum _{r=1}^{N}A_{ir}q_{r}\ ,\eqno(A1)$$
and the voltage drop across line $r=(ij)$ is
$$v_{r}=v_{(ij)}=V_{j}-V_{i}=-\sum _{k=1}^{n} A_{kr}V_{k}\ .\eqno(A2)$$

It is convenient to define a diagonal matrix $\beta $, whose diagonal matrix
elements are the
conductance  $\beta _{r}=1/\alpha _{r}$  of line $r$.
Using Ohm's law
$$q_{r}=\beta _{r}v_{r}\ ,\eqno(A3)$$
(A1) and (A2), one gets
$$p=-(A\beta A^{t})V\equiv -YV\ ,\eqno(A4)$$
where $p,V$ are $n $-dimensional vectors with components $p_{i}$ and
$V_{i}$.
{}From this, one can compute the power consumed
by the network to be
$$P=-V\.p=VYV\ .\eqno(A5)$$

Absolute voltage levels in a network are meaningless; voltages
can be shifted by a common amount without affecting the physical
contents. For example, the invariance of $p$ under this shift
can be seen in (A4) from the identity
$$\sum _{j}Y_{ij}=0\ ,\eqno(A6)$$
which in turn follows algebraically from the obvious relation
$\sum _{i}A_{ir}=0$. Being a symmetric matrix, we must also have
$\sum _{i}Y_{ij}=0$, and in (A4) this simply expresses the conservation of the
external currents:
$$\sum _{i=1}^{n} p_{i}=0\ .\eqno(A7)$$

 Because of (A6), the matrix $Y$ is singular,
so (A4) cannot be inverted to obtain
$V$ as a function of $p$. This is as it should be because
the absolute level of $V$ cannot be determined.
In order to invert (A4), a common voltage level must  be
chosen and fixed.
This will be referred to
as choosing a {\it level scheme}. One simple choice is to ground the
last vertex by setting
$V_{n} =0$
(the {\it primitive level scheme}).
The main advantage of this scheme is the easiness to solve for $V$:
(A4) can now be inverted to give
$$V_{0}=-Y_{0}^{-1}p_{0}\equiv -Z_{0}p_{0}\ ,\eqno(A8)$$
where the subscript $0$ refers to vectors and matrices made up of
the first $n-1$ components.
Incorporating $V_{n}=0$,
we can enlarge this to a $n $-dimensional relation
\b{
V&=-Z'p\ ,&(A9)\cr
Z'&=\pmatrix{Z_{0}&0\cr 0&0\cr}\ ,&(A10)}$$
where $Z'$ is the {\it impedance matrix}.
The internal currents and the power are now given in the matrix notation
to be
\b{q&=-\beta A^{t}V=\beta A^{t}Z'p\ ,&(A11)\cr
P&=-p\.V=pZ'p\ .&(A12)}$$

Another useful circuit quantity is the $N\!\times\! N$
{\it contraction matrix}
$H$ used in (2.6). Its matrix element $H_{rs}$, for $r=(ij)$ and $s=(kl)$,
is [30]
\b{H_{rs}&=G_{rs}-\beta _{r} \delta _{rs}\ ,&\cr
G_{rs}&=[\beta A^{t}Z'A \beta ]_{rs}=\beta _{r} \beta
_{s}\{Z'_{ik}-Z'_{jk}-Z'_{il}+Z'_{kl}\}
\ .&(A13)}$$

Graph theory can be used to devise  rules so that these
circuit quantities can be read off
{\it directly} from a circuit (Feynman) diagram [28--32].
The result is given in
eqs.~(2.7)--(2.11).

Using (A9) and (A11), $\alpha _{s}G_{rs}$
can be given a simple  meaning. It is the
current flowing through line $r$ in the presence of
a unit external current flowing through the two
ends of line $s$: in at vertex $l$ and out at vertex $k$.
It then follows that
\b{
\sum_{s\in {\rm loop}}\alpha _sG_{rs}&=0\ ,&(A14)\cr
\sum_{s\in {\rm loop}}\alpha _sH_{rs}&=-1\ ,&(A15)}$$
where the sum is taken over the lines $s$ around any closed loop.
(A15) is a consequence of (A14) and the first equation of (A13).
(A14) is true because the current flowing through $r$ is zero
if every vertex around the loop receives both a unit incoming and a
unit outgoing current.

An alternative derivation is to think of $\alpha _r \alpha _s G_{rs}$
as the voltage drop across line $r$ when the external unit currents
are fed through the two ends of line $s$. Since this quantity is
symmetrical in $r$ and $s$, we can always interchange $r$ and $s$ in
the interpretation. Then
$\alpha _r\sum_{s\in{\rm loop}}\alpha _sG_{rs}$ is the total voltage
drop around a loop when the unit currents are fed through the two ends
of line $r$, and is hence zero.

The impedance matrix  is not uniquely defined on account of the
$p$-conservation relation (A7). A change
$$Z'_{ij}\to Z_{ij}=Z'_{ij}+\zeta _{i}+\zeta _{j}\eqno(A16)$$
shifts the overall level of $V_{i}$ by a common amount
$-v=-\sum _{j}\zeta _{j}p_{j}$, but  leaves
the physical quantities
$P$, $v_{r}$,
$q_{r}$, and $H_{rs}$ invariant.
The transformation (A16) is thus a {\it level transformation}
that changes the level schemes; the invariance means that
the matrix $Z'$ in the physical quantities (A11)--(A13) may be replaced
by the impedance matrix $Z$ in any {\it level scheme}.
In many ways the choice of a level scheme is analogous
to a gauge choice in gauge theories, though the two
have absolutely nothing to do with each other.

Physical quantities are defined only on the {\it momentum shell}, where
constraint (A7) is satisfied. (A9), (A11), and (A12), with $Z'$ replaced
by a general $Z$, will be used to define $V,q,P$ off-shell
 as a function of all $p_i$. Off-shell
extensions are level-scheme (or $Z$) dependent, momentum conservation
is generally invalid, though Kirchhoff's voltage law,
that the total voltage drop along a closed path is zero,  still holds.
Off-shell quantities are needed in Sec.~5 and Appendix D.

A  level scheme that is particularly
symmetrical and  useful in Sec.~5 is
the {\it zero-diagonal level scheme}, or simply the {\it diagonal
scheme}, defined by setting
$$Z_{ii}=0 \quad \forall i\ .\eqno(A17)$$
Its graphical rules can be inferred from (2.8) to be
$$Z_{ij}=-\h\sum _{T_{2}^{ij}}\prod ^{\l+1}\alpha \ ,
\quad(i\not=j)\ ,\eqno(A18)$$
where $T_{2}^{ij}$ is the set of all 2-trees with vertices $i$ and
$j$ belonging to two separate 1-trees.

We pause now to indicate briefly how (2.14) is obtained from (2.1),
and why electric circuit theory becomes relevant in scattering
amplitudes [30].

Let $Q=\h kak+kb+c$ be a quadratic function of the d-dimensional
loop-momentum vectors
 $k_{a}(1\le a\le \l)$,
and $f_0(k)$ be a polynomial function of $k_{a}$. The
Euclidean-space Gaussian integral formula
$$\int(\prod_{a=1}^\l d^dk_a) f_0(k)\exp(-Q)=
\[\frac {2\pi }{\det(a)}\]^{d/2}f_0(-\partial /\partial b)\exp\(-c+\h
ba^{-1}b\)\eqno(A19)$$
can be used to evaluate (2.1), after (2.13) is used to introduce
the Schwinger parameters and a Wick rotation is
made. For subsequent manipulation,
the first crucial point to note is that the exponent on the rhs of
(A19) is the value of $-Q$ evaluated at $\partial Q/\partial k_{a}=0$,
or at $k=-a^{-1}b$. From (2.1) and
(2.13), $Q=-i\sum_r\alpha _rq_r^2$.
The $N$ internal momenta $q_r$ are  subject to
$n-1$ momentum-conservation constraints (A1), which can be solved to
express  $q_r$ as  linear combinations of the external momenta $p_i$
and the $\l=N-n+1$  loop momenta $k_a$. The resulting $Q$ is a
quadratic function of $k$ as required. Nevertheless,
the second crucial observation is that this explicit solution
of $q_r$ in terms of $k_a$ should
{\it not} be used at this stage. Instead,
$\partial Q/\partial k_{a}=0$ should be computed
using the Lagrange multiplier method, in which
constraints are incorporated into the function
$Q'=Q-2i\sum_{i=1}^nV_i(Aq-p)_i$ via the Lagrange multipliers
 $-2iV_i$. Then $\partial Q/\partial k_{a}=0$  can be replaced by
$\partial Q'/\partial q_r=0$,
which yields Ohm's law (A3) and (A2). The
momentum-conservation equation (A1) then becomes Kirchhoff's current
law, and the solution for $q$ is given by (A11). This is how circuit
quantities enter into the scattering amplitude formula (2.13), as described
in Sec.~2. $\alpha _r,q_r,p_i,P$ become respectively the resistance,
the internal current, the external current, and the power of the network.
Spacetime points become voltages [30]
and translational invariance becomes
level invariance.

Suppose the function $f_0(k)=S_0(q,p)$ is expressed as a function of
$q$ and $p$. Eq.~(A19) can then be written as
$$\int(\prod_{a=1}^\l d^dk_a) S_0(q,p)\exp(-Q)=
\[\frac {2\pi }{\det(a)}\]^{d/2}S(q,p)\exp(iP)\ ,\eqno(A20)$$
which eventually leads to (2.14). Here
   $P$ is given by (2.5) and (2.8), $\det(a)$ is essentially $\Delta $ of
(2.7).
$q$ is given by (2.9) and (A11), and $S=\sum_kS_k$ is given by the sum of
$S_0$ and its contractions, computed according to (2.6). In the
language of (A19), contraction between a pair of loop momenta $k_d$ and
$k_e$ comes from the differentiation
$(-\p/\p b_d)(-\p/\p b_e)\h ba^{-1}b=(a^{-1})_{de}$.
For more details, please consult Ref.~[30].

We turn now to
{\it differential circuit identities} needed in
Sec.~5 and Appendix D. They are
most easily derived in the primitive
level scheme, though the result is valid in any  scheme because
only level-independent quantities are involved in the formulas.

 Using (A4)--(A13), one gets
\b{
{\p \over\p \alpha _{r}}P(\alpha ,p)&=p{\p Z'\over\p \alpha _{r}}p=
(pZ'A\beta )_{r}(\beta A^{t}Z'p)_{r}=q_{r}^{2}\ ,&(A21)\cr
{\p \over\p \alpha _{s}}q_{r}(\alpha ,p)
&=-\beta _{s}\delta _{rs}q_{s}+[\beta A^{t}Z'A\beta ]_{rs}
(\beta A^{t}Z'p)_{s}=H_{rs}q_{s}\ .&(A22)}$$
These are the two basic identities from which many other identities
can be derived.

An important consequence of (A22) is that  current conservation
leads to  a corresponding contraction-matrix conservation.
If $\sum _{r}c_{r}q_{r}=0$, then it follows from (A22) that
$$\sum _{r}c_{r}H_{rs}=0 \eqno(A23)$$
for every $s$.
In particular, if $a$ is an external cubic vertex (Fig.~8),
at which momentum conservation reads $p_{a}=q_{a''}-q_{a'}$, then
$$H_{a's}=H_{a''s}\eqno(A24)$$
because $p_a$ is $\alpha _s$ independent.

The conservation of $H_{rs}$ wrt one of the two indices,
while holding the other one fixed [eq.~(A23)], together with
(A24), show that contraction matrix  elements
act like bilinear forms of loop momenta. For a $\l$-loop diagram,
there are therefore $\l(\l+1)/2$ independent $H_{rs}$'s. If necessary,
by the polarization trick,
they can all be written as linear combinations of
diagonal elements, which according to (2.10) are simply derivatives
of $\ln(\Delta )$.

The conservation equation (A23) could have been derived from the
interpretation of $\alpha _sG_{rs}$ as a current, given
immediately  below (A13).
Note that $H_{rs}$ differs from $G_{rs}$ by $-\beta _r \delta _{rs}$.
The reason why it is $H_{rs}$ but not $G_{rs}$ that is conserved
is because there are also external current injection at the
two ends of line $s$.

Writing $P=\sum_r \alpha _rq_r^2$, using (A22) and (A21), one
gets
$$\sum_{r=1}^N \alpha _rq_rH_{rs}=\sum_rv_rH_{rs}=0\ .\eqno(A25)$$
So the voltage drops form an $N$-dimensional null vector of the matrix $H$.

We can also obtain a relation for the derivatives of $H_{rs}$ by
differentiating (A22) wrt $\alpha _{t}$ and use the commutativity of the double
derivative. Then
$${\p H_{rs}(\alpha )\over\p\alpha _{t}}=H_{rt}(\alpha )H_{ts}(\alpha )\
.\eqno(A26)$$

So far the relations derived are very general. We shall now derive a set
of relations valid only at external cubic vertices $a$ (see Fig.~8).

An important property of an external vertex $a$, which can be seen
from the graphical rules (2.7)--(2.11), and (A18),
is that $\Delta (\alpha ),
Z_{ij}$ (in the diagonal level scheme) and $H_{rs}$ depends on
$\alpha _{a'}$ only through the combination
$\alpha _{a'}+\alpha _{a''}$, provided $a$ is not equal to
$i$ or $j$.  Let
\b{
\p_{a}&=-\sum _{r} A_{ar}{\p\over\p \alpha _{r}}=-{\p\over\p
\alpha _{a''}}+{\p\over\p \alpha _{a'}}\ .&(A27)}$$
Then
$$\p_{a}\Delta (\alpha )=\p_{a}H_{rs}=\p_{a}Z_{ij}=0\eqno(A28)$$
provided $i\not=a\not=j$.
Similar
equations  do not exist for internal vertices. This is the
reason why the stringlike reorganization discussed in Sec.~5 is valid
only for external vertices.

The first two of these  relations also follow from (2.10), (A24), and (A26).
The last relation is clearly  level-dependent. The fact that it is
true in the diagonal level scheme is what makes this scheme particularly
useful.

We proceed now to derive other level-dependent relations valid in the
diagonal scheme. These relations are useful in Sec.~5. Let
$$
  I_{rm}=\beta _{r}(Z_{im}-Z_{jm})\eqno(A29)
$$
be defined by (2.3) and (A11). It is the current flowing through
line $r=(ij)$ when a unit external current flows out of vertex $m$.
We will assume  $q_r$ and $p_m$ to flow in the same direction.
If not, a minus sign should accompany the appropriate quantities below.
Note that this is an off-shell interpretation, so it is level-scheme
dependent, and that current conservation is not required or assumed.

Write (A13) in the diagonal scheme as follows,
$$(Z_{il}-Z_{jl})=(Z_{ik}-Z_{jk})
-\alpha _{r}(\alpha _{s}H_{rs}+\delta _{rs})\ ,\eqno(A30)$$
where $r=(ij)$ and $s=(kl)$.
 Let $r=s$ and use (A17). Then
$$
  I_{ri}=-\beta _{r}Z_{ij}=\h (\alpha _{r}H_{rr}+1)=-I_{rj}\ .\eqno(A31)
$$
Next, suppose  $s\not= r$. Then (A30) can be written as a recursion formula
$$
  I_{rk}=I_{rl}+\alpha _{s}H_{rs}\ .\eqno(A32)
$$
Let $s=s_q=(mi_q),s_{q-1}=(i_qi_{q-1}),\cdots,s_1=(i_2i),r=s_0=(ij)$
be a continuous path leading from vertex $j$ to $m$ (see Fig.~12). Using (A32)
repeatedly, one gets
$$
  I_{rm}=\h(1+\alpha _{r}H_{rr})+\sum_{p=1}^q\alpha _{s_{p}}H_{rs_{p}}
\ .\eqno(A33)
$$

Suppose $a$ is an external cubic vertex (see Fig.~8)
and suppose $i\not=j$. Using (A28) and
(A33), one gets
$$\p_aI_{rm}=-\rho _{arm}H_{ra'}\ ,\eqno(A34)$$
where
$$\rho _{arm}=\cases{\ 0,&if $a\not=j,i,m$;\cr
-\h,&if $a=j\not=m$ or $a=i\not=m$;\cr
 \ \h,&if $a=i=m$ or $a=m=j$;\cr
\ 1,&if $a=m\not=i,j$.\cr}\eqno(A35)$$
Note that $a=j=m$ indicates the presence of a closed loop.
The relations for $a=m$ will be particularly important. Let
$$\rho _{ar}\equiv \rho _{ara}=\cases{1,&if $A_{ar}=0$;\cr
\h,& if $A_{ar}\not=0$.\cr}\eqno(A36)$$
Then
$$\p_aI_{ra}=-\rho _{ar}H_{a'r}\ .\eqno(A37)$$
This equation will be used to replace contraction of external
cubic vertices by differentiation. See Sec.~5 and Appendix D.

Next, assume $a=i\not=m$ to be an external cubic vertex and use
(A29) to compute
$$\p_iI_{rm}=
\beta _{r}(I_{rm}+\p_i Z_{im})\ .\eqno(A38)$$
Using (A35), this gives
$$\p_i Z_{im}=\h \alpha _{r}H_{rr}-I_{rm}\ ,$$
which can be written in a form independent of the label and the
postition of $i$, as long as $a\not=m$:
$$\p_a Z_{am}=\h \alpha _{a''}H_{a''a''}-I_{a''m}\ .\eqno(A39)$$
Similarly, using (A29) and (A35), one gets
$$\p_j Z_{jm}=-\h \alpha _{r}H_{rr}-I_{rm}\ ,$$
which in a postition independent form reads
$$\p_a Z_{am}=-\h \alpha _{a'}H_{a'a'}-I_{a'm}\ .\eqno(A40)$$
Adding (A39) and (A40), one deduces that as long as $a\not=m$,
$$-2\p_a Z_{am}=\h(\alpha _{a'}-\alpha _{a''})H_{a'a'}+(I_{a'm}+I_{a''m})\
.\eqno(A41)$$
It is easy to verify that this relation holds for $a=m$ as well.
Note however that relations (A39) and (A40) individually are
not valid at $a=m$.

Multiply (A41) by $p_m$ and sum over $m$. Using the
momentum-conservation equation (A7), one gets
$$-2\sum_{m=1}^n \p_a Z_{am}p_m=q_{a'}+q_{a''}\ .\eqno(A42)$$

Using the explicit form for $I_{rm}$, it is easy to obtain from
(A41) that if $b=m\not=a$ is also an external cubic vertex, then
$$\p_a\p_b Z_{ab}=H_{a'b'}\ .\eqno(A43)$$
It should be emphasized once again that these level-dependent relations
are vald only in the diagonal scheme.

\bigskip
\centerline{\bf Appendix B: Color factors}

Let $T^{a}$ be the $U(N_{c})$ generators in the fundamental representation
normalized
according to (3.3). The completeness relation dual to (3.3) is
$$\sum _{a=0}^{N_{c}^{2}-1}(T^{a})_{ij}(T^{a})_{kl}=\delta _{il}\delta _{kj}\
.\eqno(B1)$$

This can be used to sum up intermediate color indices as follows. Let
$X,Y,U,V$ each be a product of these generator matrices. Then
(B1) implies
$$\sum _{a=0}^{N_{c}^{2}-1}(XT^{a}Y)_{ij}(UT^{a}V)_{kl}=(XV)_{il}(UY)_{kj}\
,\eqno(B2)$$
which is the basis for most of the rules for a color path.  In particular,
rule 3.

When a three-gluon vertex or a ghost-gluon vertex with colors $a,b,c$ is
encountered, the color factor for the vertex is the
Lie-algebra structure constant
$$f^{abc}=-i\{\Tr(abc)-\Tr(acb)\}\ ,\eqno(B3)$$
where $a,b,c$ on the rhs stand for $T^{a},T^{b},T^{c}$, respectively.
When a four-gluon
vertex is encountered, the color factor is
$$f^{abcd}\equiv
\sum _{e}f^{abe}f^{ecd}=(-i)^{2}\{\Tr(abcd)-\Tr(bacd)-\Tr(abdc)+\Tr(badc)\}\ .
\eqno(B4)$$
In particular, this shows the decoupling between the $U(1)$ and the
$SU(N_c)$ in $U(N_c)$:
$$f^{0bc}=f^{0bcd}=0\ .\eqno(B5)$$
Eqs.~(B3) and (B4) are the basis of  rule 5 stated in Sec~3.2 for a
color path.
The predetermined order stated in the rule is determined by which
of the terms in (B3) and (B4) are chosen.

Modifications are necessary for $SU(N_{c})$ theories because it has only
$N_{c}^{2}-1$ generators. The completeness relation is now replaced by
$$\sum _{a=1}^{N_{c}^{2}-1}(T^{a})_{ij}(T^{a})_{kl}=\delta _{il}\delta
_{kj}-{1\over
N_{c}}\delta _{ij}\delta _{kl}\ .\eqno(B6)$$
An exception occurs
if somewhere in the connected gluon network a three-gluon, a four-gluon,
or a ghost-gluon vertex occurs. Then because of (B5), we may effectively
sum over all $N_{c}^{2}$ color vertices again, so the correction term on the
rhs of (B6) is no longer necessary.

As a result, color paths and color factors
of $SU(N_{c})$ are identical to those of $U(N_{c})$, except with the following
additions corresponding to the last term of (B6). Two quark lines, or
two protions of the same quark line, may be linked by a {\it single},
or more than one gluon lines. In the latter case, everything is the
same as $U(N_c)$.
In the former case, a new option
is available when
a quark-gluon vertex is encountered while travelling along a quark line.
If this new option is taken,  the path will continue {\it along the
same quark
line} passed the vertex. At the same time, the path along the other
quark line, passing through the other end of the single internal
gluon line connecting the two, must also adopt the new option
of going straight forward.
The color factor with this new option carries an extra $(-N_{c}^{-1})$
factor but would otherwise be the same as in the $U(N_c)$ case.

In a color covering where new options are taken, the single
internal gluon lines in question will never be traversed by any
color path, whereas all the other gluon lines would be traversed twice
as before.

For example, the {\it additional} color coverings
 for Fig.~1 in the case of $SU(N_{c})$
are $\{v_{1}=(p_{2}q_{2}q_{3}p_{3}),v_{2}=(p_{4}q_{2}q_{1}p_{1})\}$,
$\{v_{1}=(p_{2}q_{1}q_{4}p_{3}),v_{2}=(p_{4}q_{3}q_{4}p_{1})\}$, and
$\{v_{1}=(p_{2}q_{1}p_{1}),v_{2}=(p_{4}q_{3}p_{3})\}$. The corresponding
color factors are
$-N_{c}^{-1}({\bf 1})_{ij'}({\bf 1})_{i'j}$,
$-N_{c}^{-1}({\bf 1})_{ij'}({\bf 1})_{i'j}$, and
$+N_{c}^{-2}({\bf 1})_{ij}({\bf 1})_{i'j'}$.

\bigskip
\centerline{\bf Appendix C. Spinor helicity technique}

 The massless Dirac spinor
in the chiral representation where
$\gamma _{5}$ is diagonal can be taken to be
\b{
u_{+}(p)|_{m=0}\equiv \ket{p+}&=
\sqrt {2p^{0}}\pmatrix{\chi _{+}(\vec p)\cr 0\cr}\ , &\cr
u_{-}(p)|_{m=0}\equiv \ket{p-}&=
-\sqrt {2p^{0}}\pmatrix{0\cr \chi _{-}(\vec p)\cr}\ ,&\cr
\bar u_{+}(p)|_{m=0}\equiv \bra{p+}&=
-\sqrt {2p^{0}}\pmatrix{0&\chi _{+}^{*}(\vec p)\cr}\ , &\cr
\bar u_{-}(p)|_{m=0}\equiv \bra{p-}&=
\sqrt {2p^{0}}\pmatrix{ \chi _{-}^{*}(\vec p)&0\cr}\ ,&(C1)}$$
where the two-component helicity eigenstate
$\chi _{\lambda }(\vec p)$ satisfies
\b{
\vec \sigma \.\vec p\chi _{\lambda }(\vec p)&
=\lambda |\vec p|\chi _{\lambda }(\vec k)\ ,&\cr
\chi ^{*}_{\lambda '}(\vec p)\chi _{\lambda }(\vec p)&
=\delta _{\lambda '\lambda }\ ,&(C2)}$$
and $p$ is in the forward light cone.
The normalization adopted in
(C1) gives
$$\bk{p\lambda |\gamma ^{\mu }|p\lambda }=2p^{\mu }\ .\eqno(C3)$$
In particular, $\xi p$ is in the forward light cone if $p$ is and
if $\xi>0$, then
\b{
\ket{(\xi p) \lambda  }=\sqrt{\xi}\ket{p \lambda }\ ,&\cr
\bra{(\xi p) \lambda }=\sqrt{\xi}\bra{p \lambda }\ &(C4)}$$

Chirality conservation implies
$$\bk{p\pm |q\pm }=0\ .\eqno(C5)$$
The non-zero matrix elements are
$$\eqalignno{ \bk{pq}&=\bk{p-|q+}=2\sqrt {p^{0}q^{0}}\chi _{-}^{*}(\vec p)\chi
_{+}(\vec q)\
,&\cr
[pq]&=\bk{p+|q-}=2\sqrt {p^{0}q^{0}}\chi _{+}^{*}(\vec p)\chi _{-}(\vec q)\
.&(C6)}$$
Since $\sigma _{2}\vec\sigma \sigma _{2}=-\vec\sigma ^{*}$, it is convenient to
choose
the phase of the helicity eigenstates to satisfy
$$\sigma _{2}\chi _{\lambda }(\vec p)=i\lambda \chi _{-\lambda }^{*}(\vec p)\
.\eqno(C7)$$
It also determines the phase under charge conjugation.
This then implies
$$\chi _{\pm }^{*}(\vec p)\chi _{\mp }(\vec q)=\chi _{\pm }^{*}(\vec
p)\sigma _{2}\sigma _{2}\chi _{\mp }(\vec q)= -\chi _{\mp }^{*}(\vec q)\chi
_{\pm }(\vec p)\
,\eqno(C8)$$
or equivalently,
$$\bk{pq}=-\bk{qp}\ ,\quad  [pq]=-[qp]=sign(p\.q)\bk{qp}^{*}\ .\eqno(C9)$$
Note that both $p$ and $q$ are assumed to be in the forward light cone
so $\p\.q\ge 0$.

This phase also gives rise to the relations
\b{
\bk{p\pm |\gamma _{\mu _{1}}\cdots\gamma _{\mu _{2n+1}}|q\pm }&=
\bk{q\mp |\gamma _{\mu _{2n+1}}\cdots\gamma _{\mu _{1}}|p\mp }\ ,&(C10)\cr
\bk{p\pm |\gamma _{\mu _{1}}\cdots\gamma _{\mu _{2n}}|q\mp }&=
-\bk{q\pm |\gamma _{\mu _{2n}}\cdots\gamma _{\mu _{1}}|p\mp }\ .&(C11)}$$
Using (C3) and
$$\gamma p=|p+\rangle \langle p+|+|p-\rangle \langle p-|\ ,\eqno(C12)$$
it is easy to see that
$$\bk{p+|\gamma k|q+}=[pk]\bk{kq}\ ,\quad
\bk{p-|\gamma k|q-}=\bk{pk}[kq]\ .\eqno(C13)$$
In particular, using (C3), one gets
$$\bk{qp}[pq]=2(p\.q)\ .\eqno(C14)$$
Thus up to phases, both $\bk{pq}$ and $[pq]$ are equal to the square root
of $2p\.q$.
This equation also dictates the sign factor $sign(p\.q)$ of (C9)
when $p\.q$ takes on a negative value.

Fierz identities give the relations
\b{
\bk{AD}\bk{CB}+\bk{AC}\bk{BD}&=\bk{AB}\bk{CD}\ ,&(C15)\cr
\bk{A+|\gamma _{\mu }|B+}\bk{C-|\gamma ^{\mu }|D-}&=2[AD]\bk{CB}\ ,&(C16)\cr
\bk{A+ |\gamma _{\mu }|B+ }\bk{C+ |\gamma ^{\mu }|D+ }&=2[AC]\bk{DB}\
.&(C17)}$$
The spin factors for fermions discussed in Secs.~4.2 and 4.3
are based on eqs.~(C13), (C16), and (C17).

To order $(m/p)$, the Dirac wave functions for small but finite mass $m$
are
\b{u_{\pm }(p)&=\ket{p\pm }+{m\over 2p^0}\ket{\tilde p\mp }\ ,&\cr
\bar u_{\pm }(p)&=\bra{p\pm }+{m\over 2p^0}\bra{\tilde p\mp }\ ,&(C18)}$$
where $\tilde p=(p^0,-\vec p)$ when $p=(p^0,\vec p)$.
If $\Gamma $ is a product of an even number of $\gamma $-matrices, then
$\bk{p\pm |\Gamma |q\pm }=0$, but the limit as $m\to 0$ of the following
exists and can be computed from (C18):
$$\lim_{m\to 0}\{m^{-1}\bar u_{\pm }(p)\Gamma u_{\pm }(q)\}=
{1\over 2p^0}\bk{\tilde p\mp |\Gamma |q{\pm }}+{1\over 2q^{0}}\bk{p{\pm
}|\Gamma |\tilde q{\mp }}\ .
\eqno(C19)$$
 These relations can be used in Sec.~5.2 to calculate the matrix
elements of the transformed vertices.

The photon/gluon polarization vector with helicity $\pm 1$ can be chosen to be
$$\epsilon _{\pm }^{\mu }(p,k)=\pm {\langle p\pm |\gamma ^{\mu }|k\pm \rangle
\over
\sqrt {2}\langle k\mp |p\pm \rangle }\ ,\eqno(C20)$$
where the {\it reference
momentum} $k$ in (C18) is massless but otherwise arbitrary.  The choice of
different $k$ corresponds to the choice of a different gauge, and these
different choices are related by
$$\epsilon _{+}^{\mu }(p,k)\to \epsilon _{+}^{\mu }(p,k')-\sqrt
{2}{\bk{kk'}\over\bk{kp}\bk{k'p}}p^{\mu }\ .\eqno(C21)$$
These polarization vectors satisfy the following identities:
$$\eqalignno{
\epsilon _{\pm }^{\mu }(p,k)&=(\epsilon _{\mp }^{\mu }(p,k))^{*}\ ,&(C22)\cr
\epsilon _{\pm }(p,k)\.p&=\epsilon _{\pm }(p,k)\.k=0\ ,&(C23)\cr
\epsilon _{\pm }(p,k)\.\epsilon _{\pm }(p,k')&=0\ ,&(C24)\cr
\epsilon _{\pm }(p,k)\.\epsilon _{\mp }(p,k')&=-1\ ,&(C25)\cr
\epsilon _{\pm }(p,k)\.\epsilon _{\pm }(p',k)&=0\ ,&(C26)\cr
\epsilon _{\pm }(p,k)\.\epsilon _{\mp }(k,k')&=0\ ,&(C27)\cr
\epsilon _{+}^{\mu }(p,k)\epsilon _{-}^{\nu }(p,k)+\epsilon _{-}^{\mu
}(p,k)\epsilon _{+}^{\nu }(p,k)
&=-g^{\mu \nu }+{p^{\mu }k^{\nu }+p^{\nu }k^{\mu }\over p\.k}\ ,&(C28)\cr
\gamma \.\epsilon _{\pm }(p,k)&=\pm {\sqrt {2}\over \bk{k\mp |p\pm }} (|p\mp
\rangle \langle k\mp |+|k\pm \rangle \langle p\pm |)\ .&(C29)\cr }$$
The spin factors for photons/gluons given in Secs.~4.2 and 4.3 are
direct results of these equations.

See the last paper in Ref.~[16] for an excellent review of
the spinor helicity technique and its application to tree amplitudes.

\bigskip
\centerline{\bf Appendix D. Gauge invariance}

We discuss in this appendix local gauge invariance
in the Schwinger representation, and from there  the
stringlike formula (5.7) for scalar QED.

Let us first review how local gauge invariance
is usually proved in the momentum space.
Consider the three diagrams in Fig.~13. Each of them is meant to be
a part of a larger, but otherwise common, diagram.
Let $\epsilon _a$ be the polarization vector of the external photon at vertex
$a$. The vertex $j$ may be internal or external.
A gauge transformation produces the change
$$\delta \epsilon _{a}=\lambda _{a} p_{a}\eqno(D1)$$
for some infinitesimal gauge
parameter $\lambda _{a}$, which in turn
produces the following changes in Figs.~13(a), (b), and (c):
\b{
\delta _{a}&=\lambda _{a} e^{2} p_{a}\.[(q+p_{a})+q]{1\over
-(q+p_{a})^{2}}[(q'-p_j)+q']^{\nu }\cr
&=\lambda _{a} e^{2}[(q+p_{a})^{2}-q^{2}]{1\over
-(q+p_{a})^{2}}[2q'-p_{j}]^{\nu }&\cr
&\to -\lambda _{a} e^{2}[2q'-p_{j}]^{\nu }\ ,&\cr
\delta _{b}&=\lambda _{a} e^{2} p_{a}\.[(q'-p_{a})+q']{1\over -(q'-p_{a})^{2}}
[(q+p_{j})+q]^{\nu }&\cr
&=\lambda _{a} e^{2}[q^{'2}-(q'-p_{a})^{2}]{1\over
-(q'-p_{a})^{2}}[2q+p_{j}]^{\nu }&\cr
&\to \lambda _{a} e^{2}  [2q+p_{j}]^{\nu }\ ,&\cr
\delta _{c}&=2\lambda _{a} e^{2} p_{a}^{\nu }\ .&(D2)}$$
The arrows single out the term that cancels the propagator. The remaining
term will cancel another propagator (not shown) on the other side
of the vertex.
Thus the total local change is
$$\delta _{a}+\delta _{b}+\delta _{c}\to
-e^{2}\lambda _{a}[(2q'-p_{j})-(2q+p_{j})-2p_{a}]^{\nu }=0\ ,\eqno(D3)$$
which shows local gauge invariance at an external vertex.

In short, the proof depends on the cancellation of an adjacent
propagator from the divergence of the vertex, thereby reducing both
Figs.~13(a) and 13(b) to the seagull form 13(c), and
 the coefficients of these three identical diagrams add up to zero.

In spite of its
similarity to (2.1), this proof of local gauge invariance
 will not work for (2.14)
because of the presence of the additional terms $S_{k}$, and because
the propagator $T$ in (2.15) can no longer be canceled
by the divergence of the vertex factor. Instead, we shall have to
adopt a new technique in which Figs.~13(a) and 13(b)
are turned into the seagull
form Fig.~13(c) by {\it shrinking} the relevant propagators, rather
than by canceling them.

The success of this  depends crucially on the
 two differential circuit identities
(A21) and (A22).
Under the gauge change (D1),
the change of the vertex
$C_{a}=e\epsilon _a\. (q_{a'}+q_{a''})$ can be obtained from (A21)
and the conservation equation
$$p_a=q_{a''}-q_{a'}\eqno(D4)$$
to be
\b{\delta C_{a}&=\lambda _{a} ep_{a}\.(q_{a'}+q_{a''})
=-\lambda _{a} e\p_{a}P\ ,
&(D5)}$$
where $\p_{a}$ is defined in (5.10).
The gauge change of the
contraction of $C_{a}$ with any internal momentum $q_{r}$
is
$$\delta (C_{a}\sqcup q_{r})=\lambda _{a}
(-ie/2)(H_{a'r}+H_{a''r})p_{a}=-ie\lambda _{a}
H_{a'r}p_{a} \eqno(D6)$$
when (A24) is used, and using (A22) this is equal to
$$\lambda _{a}(ie\p_{a})q_{r}=ie\lambda _{a}
(H_{a'r}q_{a'}-H_{a''r}q_{a''})=-ie\lambda _{a}
H_{a'r}p_{a}\ .\eqno(D7)$$
This means that the momenta in $C_{a}$ and their contractions can both be
obtained from the operator $ie\p_{a}$.

Consequently, if $C_{a}$ is factored out
of the numerator factor $S_{0}$ in (2.1),
$$S_{0}(q,p)=C_{a}S'_{0}(q,p)\ ,\eqno(D8)$$
the gauge change of the amplitude
$$A=\int [D_S\alpha ] \Delta ^{-d/2} S(q,p) \exp[-i\{M-P\}]\eqno(D9)$$
under (D1) is simply
$$\delta A=\int [D_S\alpha ](ie\lambda _{a}\p_{a})\{
\Delta ^{-d/2}S'(q,p)\exp[-i(M-P)]\}\ ,\eqno(D10)$$
where $S'=\sum _{k}S'_{k}$ is the sum of $S'_{0}$ and all its contractions.

Instead of having the propagators cancelled, as in
(D2), local gauge invariance under (D1) is now accomplished
by the operator $(ie\p_{a})$, which upon integration, produce two terms
of opposite signs  respectively  with $\alpha '_{a}$ and $\alpha ''_{a}$
zero. Since the $\alpha $'s are resistances, setting them zero is equivalent
to short-circuiting that branch, which effectively turns
Figs.~13(a) and 13(b) into the seagull form of Fig.~13(c).
 The rest of the proof of local gauge invariance is similar to (D3).

Under a reparametrization transformation of the proper time
$\tau $, a function $f(\tau )$ is changed by an amount $(\delta \tau )
(/p f(\tau )/\p \tau )$. Taking the Schwinger parameters as
proper-time parameters, eq.~(D10) can be interpreted as a
special kind of proper-time
transformation of the integrand of (D9), with the vertex $a$ removed.
In that sense a gauge transformation can be regarded as a proper-time
reparametrization.
This is tentalizing because in string theories gauge invariance follows
from worldsheet reparametrization and conformal invariances. In the
infinite tension limit, where string theory becomes a gauge theory,
the worldsheet variable $\sigma $ disappears
leaving behind only the proper time $\tau $. Conformal invariance also
vanishes so  only proper-time reparametrization  is left,
nevertheless gauge invariance still holds. It is therefore tempting
to relate gauge invariance with proper-time reparametrization invariance.
Indeed such a relation exists at least in the form of eq.~(D10).

We can now repeat the same argument and compute the change of the quantity
inside the curly bracket of (D10) when another external photon $b$ undergoes a
gauge change, and so on down the line to obtain the total change to be
\b{
\delta A&=\int [D_S\alpha ](ie\prod _{a=1}^{n_{3}}\lambda _{a}\p_{a})\{
\Delta ^{-d/2}S^{int}(q,p)\exp[-i(M-P)]\}&\cr
&\equiv
\int [D_S\alpha ]\exp[ie\sum _{a=1}^{n_{3}}\lambda _{a}\p_{a}]_{ML}\{
\Delta ^{-d/2}S^{int}(q,p)\exp[-i(M-P)]\}\ ,&(D11)}$$
where $S^{int}(q,p)$ is the same internal-vertex numerator function
used in (5.7).
The subscript ML for the exponential indicates
that  only the term multilinear in all
the $\p_{a}$ should be kept when that
exponential is expanded into power series.

Eq.~(D11) has the advantage that contractions with, and among, the external
vertices  $C_{a}$  are
automatically carried out by the
operators $ie\lambda _{a}\p_{a}$,  which according to (D5) are needed
to  produce the internal momenta at
the vertices $a$ to begin with.
Now $\delta A$ is obtained from $A$ by having $\epsilon _{a}$  replaced by
$\lambda _{a}
p_{a}$. Conversely, one might think that $A$ can be obtained from $\delta A$ by
replacing $\lambda _{a} p_{a}$ with $\epsilon _{a}$, or by replacing
$[ie\lambda _{a}\p_{a}]$ with $[ie\p_{a}\epsilon _{a}\.(\p/\p p_{a})]$.
In that case, one gets
\b{
A&=\int [D_S\alpha ]\exp[ie\sum _{a=1}^{n_3}\epsilon _{a}\.D_{a}]_{ML}\{
\Delta ^{-d/2}S^{int}(q,p)\exp[-i(M-P)]\}\ ,&(D12)\cr
D^{\mu }_{a}&\equiv \p_{a}{\p\over\p p_{a\mu }}\ .&(D13)}$$
By carrying out the momentum translation implied by the first exponential
factor, one gets (5.7) (except for the $\eta_{as}$ factor in (5.11)).

The argument given above, though simple and intuitive,  is actually
full of flaws. For one thing
(D13) is meaningful only when we know how to extend the functions
it operates on off-shell (see Appendix A), but no specification has
been given in (D12) as to what level scheme to be used for such an
off-shell extension. Another point is that while it is completely
correct to obtain
$\delta A$  from $A$ by the substitution $\epsilon _a\to \lambda _ap_a$, the
inverse substitution to get from $\delta A$ to $A$ is full of uncertainties
because one would not know which $p_a$'s in $\delta A$ are to be so replaced.
The argument leading to (D12)
was presented mainly to show the closeness between the
stringlike formula and the local gauge invariance considerations.
Beyond that it should not be taken seriously.

The correct derivation makes use of (A37), (A42), and (A43).
If $a,b$ are external cubic vertices, and there are $n_3$
of those altogether, then (A42) and (A43) show that the external
cubic vertices
$\prod_{a=1}^{n_3}C_a=e\prod_{a=1}^{n_3}
\epsilon _a\.(q_{a'}+q_{a''})$, as well as their mutual contractions
$C_a\sqcup C_b=-2ie^2
\epsilon _a\.\epsilon _bH_{a'b'}$, can all be obtained from
$\exp(iP)=\exp(i\sum_{i,j}p_i\.p_jZ_{ij})$ by shifting $p_a$
to $p_a+ie\epsilon _a\p_a$, and taking the terms multilinear (ML) in the
$\epsilon _a$'s.
This proves (5.7) when $S_0^{int}=1$.
When $S_0^{int}(q,p)\not=1$, contractions of pairs of $q$'s within
$S_0^{int}$, as well as contractions of
those in $S_0^{int}$ with those in $C_a$
are necessary. The former turns $S_0^{int}$ into $S^{int}(q,p)$,
and the latter, $C_a\sqcup q_r=-ie\epsilon _aH_{a'r}$,
according to (A37) can be accomplished
by  shifting $p_a$ within $q_r$ to $p_a+ie\rho _{ar}^{-1}\epsilon _a\p_a$,
as stated in (5.7) and (5.9).

So far we have considered gauge invariance for  external photon lines.
The proof of local gauge invariance for  internal photon lines is similar but
more complicated. The main difference is that momentum conservation
and (A23) have to be used more extensively. Here is how it goes.

The vertex $i$ in Fig.~14 is meant to be the same as
the vertex $a$ in Figs.~13a, except that it is now internal.
Being internal, the photon line with momenta $p_{i}$ must
connect to another charged  line at a vertex $k$, as shown.
Factor out the explicit factors connecting
vertices $i$ and $k$, we can write the numerator
function
$S_{0}$ of (2.1) in the Feynman gauge as
$$S_{0}=e^2[(q_{i''}+q_{i'})\.(q_{k''}+q_{k'})]
S_{0}'\ .\eqno(D14)$$
Using (5.5), a gauge change of the photon joining
$i$ and $k$ would lead to a change
$$\delta S_{0}=(\delta \xi
)[p_{i}\.(q_{i''}+q_{i'})][p_{i}\.(q_{k''}+q_{k'})]p_{i}^{-2}
S_{0}'\ .\eqno(D15)$$
Strictly speaking, the denominator $p_i^{-2}$ above should not
appear in $\delta S_0$.
When we go over to the parametric representation (2.14), this
 factor  can be replaced by a factor $i\alpha _{i}$, so we
can effectively ignore this factor from now on.

Defining
$$\p_{i}=\p/\p\alpha _{i'}-\p/\p\alpha _{i''}\ ,\quad
\p_{k}=\p/\p\alpha _{k'}-\p/\p\alpha _{k''}\ ,\eqno(D16)$$
then the claim is that the change of $A$ in (2.14) is given by
$$\delta
A=(\delta \xi )\int [D_S\alpha](ie\p_{i})(-ie\p_{k})
\{i\alpha _{i}\Delta ^{-d/2}(\alpha )
S'(q,p)\exp[-i(M-P)]\}\ ,\eqno(D17)$$
where $S'(q,p)$ is the sum of $S'_{0}$ and its contractions $S'_{k}$.
The proof follows.

Using (A21) and momentum conservation, it follows like before that
$$\p_{i}P=-p_{i}\.(q_{i''}+q_{i'})\ ,\quad \p_{k}P=p_{i}\.(q_{k''}+q_{k'})\ .
\eqno(D18)$$
To show that these differential operators $\p_{i}$ and $\p_{k}$ also generates
contractions correctly, let us take an internal momentum $q_{r}$  and
compute its contraction with $p_{i}\.(q_{i''}+q_{i'})$:
$$p_{i}\.(q_{i''}+q_{i'})\sqcup
q_{r}=-(i/2)[H_{ir}(q_{i''}+q_{i'})+(H_{i''r}+H_{i'r})p_{i}]\ .\eqno(D19)$$
On the other hand, using (A22),
$$(i\p_{i})q_{r}=i[H_{i'r}q_{i'}-H_{i''r}q_{i''}]
=(i/2)[(H_{i'r}-H_{i''r})(q_{i'}+q_{i''})+(H_{i'r}+H_{i''r})(q_{i'}-q_{i''})]
\ .\eqno(D20)$$
Using momentum conservation and contraction-function conservation (A23),
eqs.~(D20) and (D19) are seen to be identical. So $(i\p_{i})$ does produce
the correct contraction with any internal momentum $q_{r}$, and similarly the
same is true for $(-i\p_{k})$.

The remaining item to consider is the effect of $(i\p_{i})$ and
$(-i\p_{k})$ on $\Delta ^{-d/2}$ and where that comes from. According to
(2.10),
$$(i\p_{i})\Delta ^{-d/2}=(id/2)\Delta ^{-d/2}(H_{i'i'}-H_{i''i''})\
.\eqno(D21)$$
The factor $(id/2)\Delta ^{-d/2}(H_{i'i'}-H_{i''i''})$
is what one would get from the contraction within the factor
$-p_{i}\.(q_{i''}+q_{i'})$,
$$-p_{i}\sqcup(q_{i''}+q_{i'})=(i/2)g_{\mu }^{\mu }(H_{ii'}+H_{ii''})\
,\eqno(D22)$$
when $d=g_{\mu }^{\mu }$ and (A23)  are used.
Other contractions can also obtained by differentiations. Thus
eq.~(D17) is   valid. The rest of the proof is very
similar to the situation in momentum space.

Once again eq.~(D17) supports the interpretation of gauge invariance
to be a proper-time reparametrization invariance.

\bigskip
\centerline{\bf Appendix E. Gordon Identity}

The off-shell Gordon identity used in Sec.~5.2 was proven in Ref.~[27].
An alternative, diagramatic, proof will be given in this appendix.

The central identity to use is
$$e[\gamma A(m+\gamma q')-(m-\gamma q'')\gamma A]=C^\mu A_\mu +S^\mu A_\mu \
,\eqno(E1)$$
where
\b{C^\mu&=e(q'+q'')^\mu\ ,&\cr
S^\mu&=-ie\sigma ^{\mu \nu}(q'-q'')_{\nu }\ .&(E2)}$$
Apply this to Fig.~13a, now considered as a part of a {\it spinor} QED
diagram, with $q''\equiv q+p_a=q'-p_{j}$ and spinor vertices $F^\mu A_ \mu
=e \gamma ^\mu  A_ \mu $:
\b{
e^{2}\gamma A_a[m-\gamma q'']^{-1}\gamma A_j[m-\gamma q']^{-1}
&=e\gamma A_a\{e\gamma A_j+[m-\gamma q'']^{-1}[C\. A_{j}+S\. A_{j}]\}
(m^2-q^{'2})^{-1}\ .&\cr
&&(E3)}$$
A similar result is obtained for Fig.~12b. In this way the scalar
propagator, the cubic scalar vertex $C$ and the magnetic-moment vertex
$S$ emerge. The seagull scalar vertex is obtained by combining the first
terms of Fig.~12a and 12b to get $e^{2}\{\gamma A_{a},
\gamma A_{j}\}=2e^2g^{\mu \nu }A_{a\mu }A_{j\nu }
\to Q^{\mu \nu }A_{\mu }A_{\nu }$.

At a vertex next to an incoming electron line with momentum $q'$,
the vertex factor $e\gamma A$ can be replaced by the factor $e\gamma A(m+\gamma
q')/2m$,
to enable (E1) to be used to derive (E2) and (E3). This accounts
for the extra factor $(2m)^{-1}$ per incoming electron. No factor
is needed for an outgoing electron of momentum $q''$ because the second
term on the lhs of (E1), when operating on the
outgoing electron, is zero and can be added in freely.

\vfill\eject
\centerline{\bf References}
\bigskip

\bigskip

\def\i#1{\item{[#1]}}
\def\npb#1{{\it Nucl.~Phys. }{\bf B#1}}
\def\plb#1{{\it Phys.~Lett. }{\bf #1B}}
\def\prl#1{{\it Phys.~Rev.~Lett. }{\bf B#1}}
\def\prd#1{{\it Phys.~Rev. }{\bf D#1}}
\def\pr#1{{\it Phys.~Rep. }{\bf #1}}
\def\ibid#1{{\it ibid.} {\bf #1}}
\parskip0pt
\i 1 F.A. Berends, R. Kleiss, P. De Causmaecker, R. Gastmans,
W. Troost, and T.T. Wu, \plb{103} (1981), 124;
\npb{206} (1982), 61;
\ibid{239} (1984), 382; \ibid{239} (1984), 395; \ibid{264} (1986), 243;
\ibid{264} (1986), 265.
\i 2  P. De Causmaecker, R. Gastmans,
W. Troost, and T.T. Wu, \plb{105} (1981), 215; \npb{206} (1982), 53.
\i{3} Z. Xu, D.-H. Zhang, and L. Chang, Tsinghua University Preprints,
Beijing, China, TUTP-84/4, TUTP-84/5, TUTP-84/6; \npb{291} (1987), 392.
\i 4 F.A. Berends and W.T. Giele, \npb{294} (1987), 700; \ibid{306} (1988),
759; \ibid{313} (1989), 595.
\i 5 F.A. Berends, W.T. Giele, and H. Kuijf, \plb{211} (1988), 91; \ibid{232}
(1989), 266; \npb{321}
(1989), 39;\ibid{333} (1990), 120.
\i 6 J. Gunion and J. Kalinowski, \prd{34} (1986), 2119.
\i {7} J. Gunion and Z. Kunszt, \plb{159} (1985), 167; \ibid{161} (1985), 333;
\ibid{176} (1986), 477.
\i{8} K. Hagiwara and D. Zeppenfeld, \npb{313} (1989), 560.
\i{9} R. Kleiss and H. Kuijf, \npb{312} (1989), 616.
\i{10} R. Kleiss and W.J. Stirling, \npb{262} (1985), 235.
\i{11} D. Kosower, \npb{315} (1989), 391; \ibid{335} (1990), 23.
\i{12} J.G. K\"orner and P. Sieben, \npb{363} (1991), 65.
\i{13} Z. Kunzt, \npb{271} (1986), 333.
\i{14} M. Mangano, \npb{309} (1988), 461.
\i{15} M. Mangano, S. Parke, and Z. Xu, \npb{298} (1988), 653.
\i{16} M. Mangano and S.J. Parke, \npb{299} (1988), 673; \prd{39} (1989), 758;
       \pr{200} (1991), 301.
\i{17} Z. Bern and D.K. Kosower, \npb{362} (1991), 289.
\i{18} S. Parke and T. Taylor, \plb{157} (1985), 81; \npb{269} (1986), 410;
\prl{56} (1986), 2459; \prd{35} (1987), 313.
\i{19} C. Dunn and T.-M. Yan, \npb{352} (1989), 402.
\i{20} G. Mahlon and T.-M. Yan, \prd{47} (1993), 1776;
G. Mahlon, T.-M. Yan, and C. Dunn, \prd{47} (1993), 1120.
\i{21} G. Mahlon, Cornell preprint CLNS 92/1154; Cornell University Ph.D.
thesis.
\i{22} Z. Bern and D.K. Kosower, \prl{66} (1991), 1669; \npb{379} (1992), 451;
 preprint Fermilab-Conf-91/71-T.
\i{23} Z. Bern and D.C. Dunbar, \npb{379} (1992), 562.
\i{24} Z. Bern, L. Dixon, and D.A. Kosower, \prl {70} (1993), 2677.
\i{25} M. Strassler, \npb{385} (1992), 145; SLAC preprint SLAC-PUB-5978.
\i{26} C.S. Lam,  \npb{397} (1993), 143.
\i{27} C.S. Lam, \prd{48} (1993), 873.
\i{28} J.D. Bjorken, Standford Ph.D. thesis (1958);
 J.D. Bjorken and S.D. Drell, `Relativistic Quantum Fields'
(McGraw-Hill, 1965).
\i{29} K. Symanzik, {\it Progress of Theoretical Physics} {\bf 20}
 (1958), 690.
\i{30} C.S. Lam and J.P. Lebrun, {\it Nuovo Cimento} {\bf 59A} (1969), 397.
\i{31} N. Nakanishi, `Graph Theory and Feynman Integrals' (Gordon and
Breach, 1971).
\i{32} P. Cvitanovi\'c and T. Kinoshita, \prd{10} (1974), 3978;
\prd{10} (1974), 3991.
\i{33} C.S. Lam, {\it Nuovo Cimento} {\bf 59A} (1969), 422.
\i{34} J. Paton and Chan Hong-Mo, \npb{10} (1969), 519.
\i{35} M. Veltman, \npb{319} (1989), 253.
\i{36} D. Kosower, \plb{254} (1991), 439.
\i{37} L.F. Abbott, \npb{185} (1981), 189; L.F. Abott, M.T. Grisaru, and
R.K. Schaefer, \npb{229} (1983), 372.

\vfill\eject

\centerline{\bf Figure Captions}

\bigskip\bigskip

\item{Fig.~1.} A QCD  diagram with internal momenta $q_r$ and
Feynman (Schwinger) parameters $\alpha_r$. Each external fermion
line is lablelled by (momentum, color, helicity).
\item{Fig.~2.} A QCD diagram with internal momenta $q_r$ and external
momenta $p_i$. Each external line is labelled by (momentum, color).
\item{Fig.~3.} A QED or QCD diagram with internal momenta $q_r$
and external momenta $p_i$. The external line labels are (momentum,
helicity).
\item{Fig.~4.} A QED or QCD diagram with internal momenta $q_r$
and external momenta $p_i$. The external line labels are (momentum,
helicity).
\item{Fig.~5.} Modified diagram of Fig.~4 used to compute spinor paths
and spin factors. $k_i$ are the reference momentum of the gluon whose
actual momentum is $p_i$.
\item{Fig.~6.} Color-oriented  QCD vertices in the background gauge used in
color and spin paths.
The
quark, gluon, and ghost lines are respectively represented by
a thick solid line, a thin solid line, and a dashed line.
External
gluon lines are indicated by a circled `A'. Color is
arranged in clockwise order, with color factors given by the trace of
products of generators in that order for gluons and ghosts. For the
quark-gluon vertex in diagram (t), the color factor is $T^a$.
 All
momenta are assumed to be outgoing, with the spin factors written
below the diagrams. The Lorentz index is that of the dislodged line,
the presence of the coupling constant $g$ or $g^2$ is understood.
Because of the unusual normalization (3.3), this coupling constant
$g$ is $1/\r2$ of the usual one.
For example, if the (momentum, Lorentz index, color index)
labels of line 1 is $(q_1,\alpha,a)$, of line 2 is $(q_2,\beta,b)$, etc.,
the total vertex factor for Fig.~6(a) is $g\Tr(T^aT^bT^c)(q_2-q_3)_\alpha$,
and for Fig.~6(e) is $+2g^2\Tr(T^aT^bT^cT^d)g_{\alpha\gamma}g_{\beta\delta}$,
and for Fig.~6(t) is $gT^a\gamma_\alpha$.
\item{Fig.~7.} A spinor path for Fig.~2, where $k_i$ is the reference
momentum for an external gluon whose actual momentum is $p_i$.
Each external line is labelled by (momentum, helicity).
\item{Fig.~8.} A cubic electromagnetic vertex for scalar QED.
The solid and dashed lines are respectively the charged scalar and
photon lines.
\item{Fig.~9.} A quartic (seagull) vertex for scalar QED.
The solid and dashed lines are respectively the charged scalar and
photon lines.
\item{Fig.10.} $m$-loop vector paths for a pure QCD diagram. Here $m=6$.
\item{Fig.11.} Some other possible vector paths for the same $m$-loop
as Fig.~10.
\item{Fig.12.} A figure used to demonstrate eq.~(A33).
\item{Fig.13.} A portion of a scalar QED diagram used to demonstrate
local gauge invariance of an external photon line.
\item{Fig.14.}  A portion of a scalar QED diagram used to demonstrate
local gauge invariance of an internal photon line.

\end